\documentclass[nofootinbib, pra, amsmath, amssymb, twocolumn]{revtex4-1}

\usepackage{graphicx}
\usepackage{dcolumn}
\usepackage{bm}

\usepackage{inputenc}
\usepackage{fontenc}
\usepackage{etoolbox}
\usepackage{hyperref}

\usepackage{subfigure}
\usepackage{verbatim}
\usepackage{xcolor}
\usepackage{tikz}
\usetikzlibrary{decorations.pathreplacing}

\begin{document}

\title{Algorithm to extract direction in 2D discrete distributions and a continuous Frobenius norm}

\author{Jeffrey G. Yepez}
\email{jgyepez@hawaii.edu}
\author{Jackson D. Seligman}
\author{Max A. A. Dornfest}
\author{Brian C. Crow}
\author{John G. Learned}
\affiliation{Department of Physics and Astronomy, University of Hawai`i at M\={a}noa, Honolulu, Hawai`i 96822}
\author{Viacheslav A. Li}
\affiliation{Lawrence Livermore National Laboratory, Livermore, California 94550}

\date{16 January 2026}

\begin{abstract}
In this study, we present a novel algorithm for determining directionality in 2D distributions of discrete data. We compare a reference dataset with a known direction to a measured dataset with an unknown direction by the Frobenius norm of the difference (FND) to find the unknown direction. To generalize this concept, we develop a continuous Frobenius norm of the difference (CFND) as a continuous analog of the FND and derive its analytical expression. By relating fitted and normalized 2D Gaussian distributions, we show that the CFND approximates the FND, and we validate this relationship with computer simulations. We find that a first-order approximation of the CFND between two similar Gaussian distributions takes the form of an absolute sine function, offering a simple analytical form with potential for specialized applications in segmented inverse beta decay (IBD) neutrino detectors, astronomy, machine learning, and more. Our methodology consists of modeling a 2D Gaussian distribution, binning the data into a histogram, and encoding it as a square matrix. Rotating this matrix around its geometric center and comparing it to a measured dataset using the FND gives us rotational data that we fit with an absolute sine function. The location of the minimum of this fit is the angle closest to the true angle of the direction in the measured dataset. We present the derivation and discuss initial applications of the CFND in our novel algorithm, demonstrating its success in approximating directionality in 2D distributions.
\end{abstract}

\maketitle

\section{Introduction}
\label{sec:introduction}
Extracting directional information from discrete data is a fundamental problem that spans many fields including physics, astronomy, and machine learning. Our motivation arises from research in reactor-antineutrino detection, specifically focusing on inverse beta decay (IBD) events as our source of directional information \cite{PhysRevApplied.22.054030}. Despite its physics origin, the core challenge is how to extract directionality and quantify confidence from a 2D dataset, which is a broadly relevant problem. This paper does not address the angular uncertainty of the direction of the dataset---this topic is reserved for another paper. Rather, this paper focuses on introducing a general framework we call the continuous Frobenius norm of the difference (CFND), which is presented in this paper as the continuous analog of the Frobenius norm of the difference (FND). We believe that the FND has potential for widespread interdisciplinary application, as presented in this report and discussed further in the final remarks in Section~\ref{sec:conclusion}.

To communicate our findings to a broad scientific audience, we begin by defining a 2D histogram of a bivariate Gaussian distribution as a matrix. In Section~\ref{sec:directionalgorithm}, we introduce the CFND direction algorithm, which leverages matrix similarity quantified through the Frobenius norm as a measure of direction. Then, in Section~\ref{subsec:simulations}, we demonstrate this method in a set of computer simulations. The Frobenius norm is chosen as the foundation of this study due to its well-established and simple utility in comparing matrices \cite{higham1988matrix}.

Our conclusion that the CFND may be a novel contribution stems from a comparison with prior work in this subject area. For instance, McElroy and Roy \cite{10.1111/rssb.12480} utilize the integral of the Frobenius norm but do not express its continuous analog as an integral. Likewise, the 2014 paper by Townsend and Trefethen \cite{townsendtrefethen} explores continuous analogs of matrix factorizations and even touches on the Frobenius norm, yet does not formalize its continuous form as a mathematical entity. They discuss continuous analogs by comparing discrete matrices with what are referred to as ``quasimatrices" and ``cmatrices," which are continuous in one and two dimensions, respectively. These are, in essence, continuous representations of matrices in which discrete entries are replaced by continuous distribution functions. In this work, we create a continuous representation of the Frobenius norm and apply it directly in a novel algorithm to determine the directionality of a 2D dataset.

\subsection{Organization}
\label{subsec:organization}
In Section \ref{sec:prerequisites}, we review concepts that should be made familiar to a physicist, mathematician, or engineer reading this paper. We go over the Frobenius norm, the normal distribution, and discuss the fundamentals histograms. We also discuss the relationship between the histogram's continuous fitted distribution function and the normal distribution---the process of normalizing a fitted distribution function.

In Section~\ref{sec:directionalgorithm}, we introduce the novel direction algorithm for 2D discrete distributions. This includes a definition of the Continuous Frobenius norm of the difference as a double integral and a derivation of its analytical form. A first-order approximation of this function is derived using a first-order Taylor series expansion to model application to neutron captures in a segmented neutrino detector. The analytical function is found for both the Gaussian and Cauchy distribution. We also discuss our algorithm methodology and some simulations that were carried out to explore the CFND relationship.

In Section~\ref{sec:conclusion}, we have a short discussion on our results and conclude this report. We give our final remarks and propose future outlooks for the project.

\section{Prerequisites}
\label{sec:prerequisites}
Here we review key concepts that are used in the mathematics presented this report. These primarily include the Frobenius norm for matrices and the normal distribution or Gaussian distribution as attributed to the German physicist Johann Carl Friedrich Gauss. We also review basic histogram concepts and discuss the relationship between a histogram and its continuous fit function. We also discuss the process of normalization of a fitted Gaussian distribution function and mathematically express the correlation of the parameters of histogram data and parameters of the fit. Utilizing this idea, we demonstrate a unique method to normalize a continuous distribution function with the discrete parameters of the histogram. This relationship is crucial in the final steps of the derivations shown in this report.

\subsection{Frobenius norm}
\label{subsec:frobeniusnorm}
The Frobenius norm is a type of matrix norm, that is a way to put a single number on a matrix. It is attributed to the German mathematician Ferdinand Georg Frobenius~\cite{frobenius1878}. It provides a measure of the magnitude of a matrix, similar to how a vector norm measures the length of a vector. In fact, the Frobenius norm is a specific case of what is known as the $p$-norm, which is a type of entry-wise norm. These norms treat an $m\times n$ matrix as a vector of size $m \cdot n$. The general $p$-norm for $p\geq 1$ is given by
\begin{align}
\|M\|_{p,p}=\|\mathrm {vec} (M)\|_{p}=\left(\sum _{i=1}^{m}\sum _{j=1}^{n}|M_{ij}|^{p}\right)^{1/p}.
\end{align}
The special case $p = 2$ is the Frobenius norm or the Hilbert–Schmidt norm, which is
\begin{align}
\label{eq:frobeniusnorm}
\|M\|_F=\sqrt{\sum_{i}\sum_{j}|M_{ij}|^2}.
\end{align}
The Frobenius norm of the difference is very similar to the well known method to determine how close two vectors are---the distance formula between two points, i.e. $ \|\bm r - \bm r'\|=\sqrt{\sum_i (r_i-r_i^{'})^2}$.
There is a striking similarity for the Frobenius norm as a measure of the length of a matrix, or in the case of the FND the distance between two matrices. The Frobenius norm is discussed in more detail in modern matrix analysis texts such as~\cite{golub1996matrix}.

\subsection{Normal distribution}
\label{subsec:normaldistribution}
In this study, we chose as part of our methodology to employ the Gaussian distribution or normal distribution in our theories. Although the mathematics in this report is generalized to any distribution function, the Gaussian distribution was selected because it is arguably one of the most profound distributions in physics, if not the most. It appears in numerous applications in experimental physics. What we widely know today as the normal distribution was invented in the 19th century by Gauss in his work entitled {\it On the Theory of the Motion of Celestial Bodies} \cite{gaussoriginal}. In Gauss's work, the first form of the normal distribution was introduced within a mere side note. The distribution was derived for the purpose of estimating uncertainty in an experimental quantity measured any number of times. In Gauss's original notation, he wrote the following equation as the final product of his derivation:
\begin{equation}
\varphi \Delta = \frac{h}{\sqrt{\pi}} \,e^{-hh\Delta\Delta},
\end{equation}
where $\varphi$ is what we today call the probability density function, $h$ is a parameter that corresponds to the width of the distribution and is therefore related to the uncertainty in the measured quantity (from Gauss's original translated text ``\dots the constant $h$ can be considered as the measure of precision of the observations."), and $\Delta$ is the dependent variable or the measured quantity.

Before going into any detail about the direction algorithm presented in this paper or the definition of the proposed mathematical constructs, it is crucial to review the 1D and 2D normal distributions. Although we discuss the generalization to the 3D case, it is not a primary focus of this report as the novel direct algorithm is described using 2D distributions and matrices. The main topic of discussion in this paper will therefore be 2D distribution functions and specifically the bivariate normal distribution.

\begin{figure}[h!]
\centering
\includegraphics[width=\columnwidth]{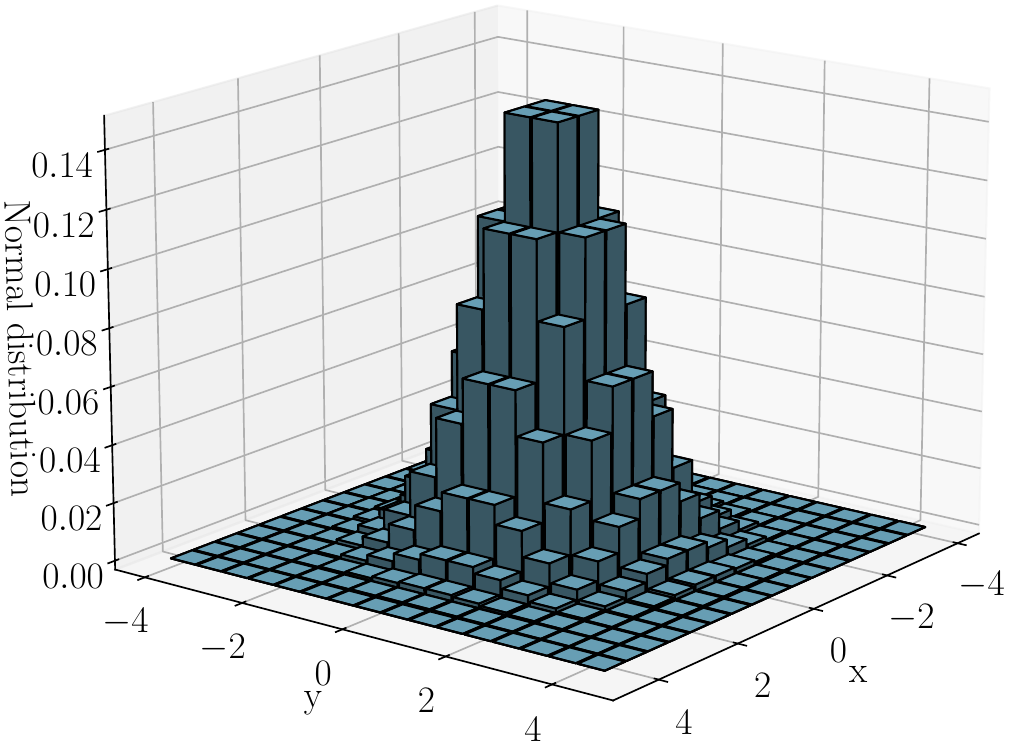}
\caption{\label{fig:legohist} 2D Gaussian histogram with standard deviation $\sigma_x=\sigma_y=1$ and centroid $\bm \mu=(0,0)$. The grid size is $16\times 16$ and the bin width $\Delta x=1/2$.}
\end{figure}

The form of the 1D normal distribution is very familiar in physics and mathematics communities. The analytical form of the normal normal distribution $N(x)$ is given by
\begin{align}
N(x)=\frac{1}{\sigma\sqrt{2\pi}}e^{-\frac{(x-\mu)^2}{2 \sigma^2}},
\end{align}
where $\sigma$ is the standard deviation, $\mu$ is the centroid or offset, and $x$ is the random variable. This distribution is shown in Figure~\ref{fig:gaussfit1d}. Conceptually, the 2D normal distribution may be thought of as two of these distributions multiplied together. The analytical form of the bivariate normal distribution is
\begin{align}
\label{eq:bivariatenormaldistribution}
N(\bm r)=\frac{1}{2\pi\sigma^2}\exp{\left(-\frac{\|{\boldsymbol r} - \boldsymbol \mu \|^2}{2\sigma^2}\right)},
\end{align}
where $\sigma$ is the variable-symmetric standard deviation. An example of a symmetrical 2D normal distribution is shown below in Figure~\ref{fig:2dnormaldist}, for standard deviation $\sigma_x=\sigma_y=1$ and centroid $\bm \mu=(0,0)$ as well as the corresponding 2D histogram in Figure~\ref{fig:legohist}. The random vector $\bm r$ and centroid $\bm \mu$ are the vectors
\begin{align}
\label{eq:gaussdistelems}
{\boldsymbol r}=
\begin{pmatrix}
    x \\ y
\end{pmatrix}
\quad \text{and} \quad
{\boldsymbol \mu}=
\begin{pmatrix}
    \mu_x \\ \mu_y
\end{pmatrix},
\end{align}
where $x$ and $y$ are the random variables whose distribution centroids are $\mu_x$ and $\mu_y$, respectively. This distribution is said to be symmetrical, that is the correlation coefficient that described the correlation between the two random variables $x$ and $y$ is $\rho=0$ and the projected standard deviations $\sigma_x=\sigma_y\equiv\sigma$.
  
In this paper, we later use the symmetrical bivariate distribution in the presentation of the direction algorithm. These equations will be referenced within the remainder of the report. To serve the interest of the reader, the normal distribution may be generalized to $k$-th order \cite{doi:10.1142/6096}, whose analytical form is given by
\begin{align}
\label{eq:generalizednormaldistribution}
N(\bm r)
=
\frac{\exp{\left(-\frac{1}{2}(\bm r - \bm \mu)^T\bm \Sigma^{-1}(\bm r - \bm \mu)\right)}}{\sqrt{(2\pi)^k|\bm \Sigma|}},
\end{align}
which is generalized to $k$ dimensions. The quantity $\bm \Sigma$ is called the covariance matrix, which in the general 2D case is
\begin{align}
\bm \Sigma = \begin{pmatrix}
 \sigma_x^2& \rho\sigma_x\sigma_y\\
 \rho\sigma_x\sigma_y& \sigma_y^2\\
\end{pmatrix}.
\end{align}
Upon simplifying (\ref{eq:generalizednormaldistribution}) for the symmetrical bivariate case where the order $k=2$, correlation coefficient $\rho=0$, and the standard deviations $\sigma_x$ and $\sigma_y$ are equal, we obtain the bivariate form of the normal distribution in (\ref{eq:bivariatenormaldistribution}).

\subsection{Fitted-normal distribution relation}
\label{subsec:fittednormalrelation}
\begin{figure}
\centering
\includegraphics[width=\columnwidth]{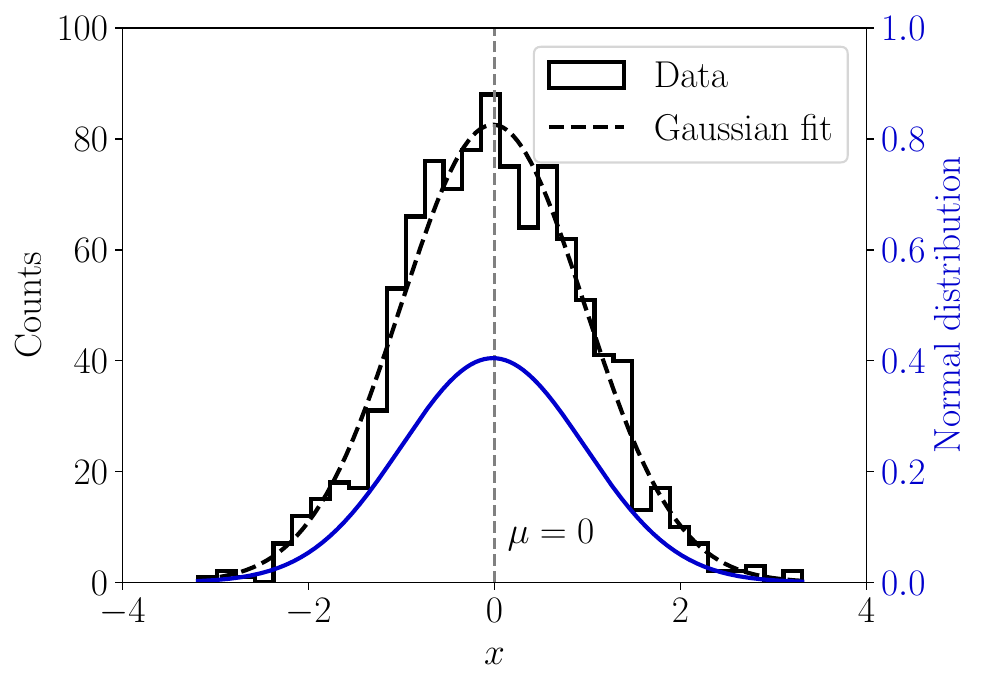}
\caption{\label{fig:gaussfit1d} Example of a Gaussian fit to a histogram of Gaussian data with standard deviation $\sigma=1$ and centroid $\mu=0$.}
\end{figure}

A common task in all fields of science is fitting Gaussian data or a Gaussian-shaped histogram. Various numerical and analytical algorithms exist that perform such a task---finding a fit to Gaussian histogram data. The method of fitting is an important and interesting topic but is not the focus of this paper. Instead we point our attention to the relationship between the fitted distribution function and its corresponding normal distribution that has the same standard deviation and centroid but normalized such that its integral over all space equals one. An example of a least squares fit to a Gaussian histogram is shown in Figure~\ref{fig:gaussfit1d}. The corresponding normalized distribution function for this fitted distribution is shown in blue in the same figure. These distributions have the same $\sigma$ and $\mu$, but the normalized distribution is scaled to integrate to one. We may normalize a distribution by simply adjusting its amplitude; changing the leading coefficient of the distribution such that the integral of the normalized distribution is equal to one. In the case of the 1D Gaussian distribution the normalizing leading coefficient is $1/\sigma \sqrt{2\pi}$. However, it is also possible to express the normalization factor of the fit function to a histogram in terms of the number of points binned in the histogram $n$ and the bin width $\Delta x$. In practice, the bin width is fixed by the experimental setup, so this relationship may be important for calibrating discrete measurement devices such as segmented neutrino detectors or CCD cameras.

\begin{figure}[h!]
\centering
\includegraphics[width=\columnwidth]{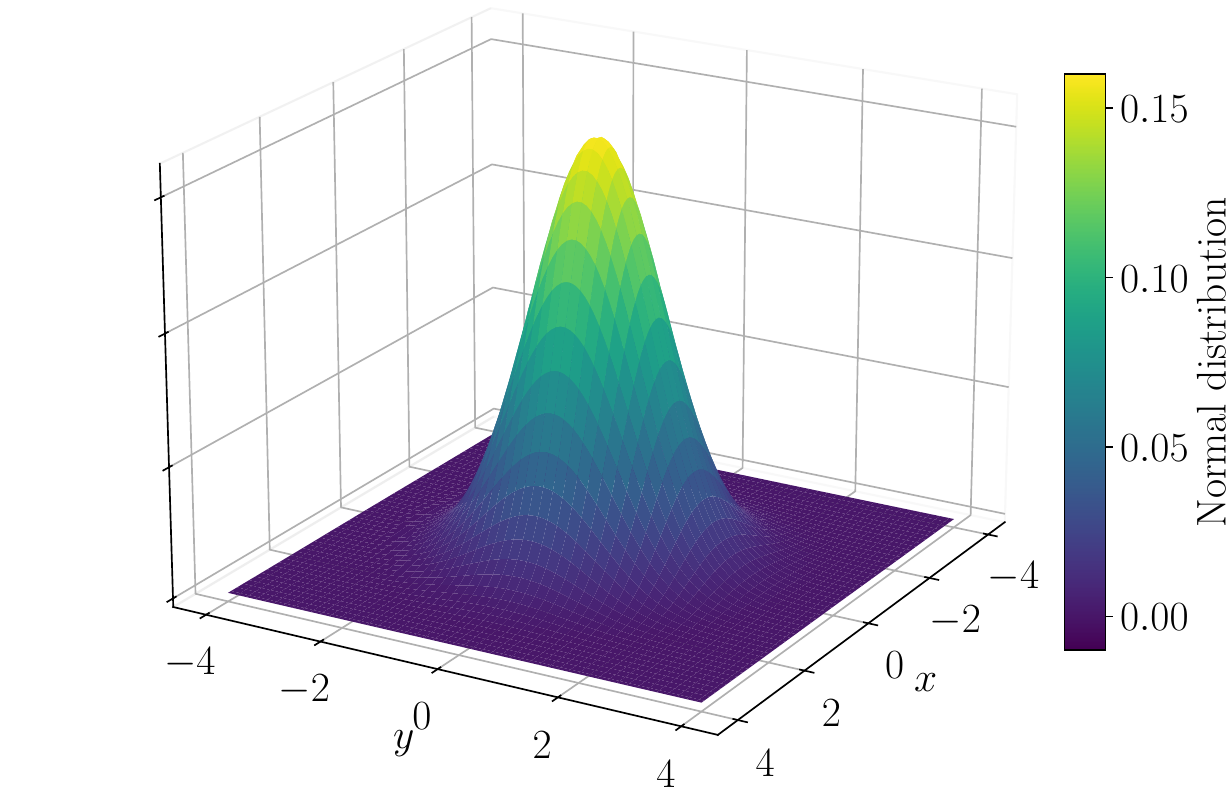}
\caption{\label{fig:2dnormaldist} Bivariate normal distribution with standard deviation $\sigma_x=\sigma_y=1$ and centroid $\bm \mu=(0,0)$.}
\end{figure}

A histogram is generally a discrete representation of a dataset that is produced by a set of repeated measurements of a random variable in a system that is governed by the physical laws of nature. A histogram is generally understood as an approximation of the true function represented by fitting the data with a distribution that represents the physics going on in the histogram, i.e. the Gaussian distribution for statistical systems. As the number of events recorded in the histogram increases $n\rightarrow\infty$ and the bin width of the histogram decreases $\Delta x\rightarrow 0$, the distinctly discrete histogram follows more closely to a continuous normal distribution.

For the remainder of our paper we shall refer to the fitted distribution as $F(\bm r)$ and the corresponding normal distribution $N(\bm r)$, that is the normal distribution will have a different amplitude but the same centroid $\bm \mu$ and covariance matrix $\bm \Sigma$ as the fitted distribution. We may derive a concrete relationship between the fitted distribution $F(\bm r)$ and the corresponding normal distribution $N(\bm r)$ that is fully dependent on the bin width and the number of events recorded in the histogram. 

For the 2D case, we have a 2D histogram of Gaussian-shaped data distributed in two random variables $x$ and $y$. We may represent 2D histogram data by a matrix we shall denote by $M$. For this explanation, let us imagine a matrix $M$ that has components $M_{ij}$, where $i$ and $j$ are the indices of the 2D array. The matrix $M$ is a full mathematical representation of the discrete histogram data. That is, the bins of the histograms have amplitude $M_{ij}$, that together form the 2D distribution. Let us suppose that there are $n$ events that are recorded in the histogram, and since $M_{ij}$ is the amplitude or counts of the elemental bin of the histogram such that the total number of events $n$ that are binned in the whole array is given by the sum of all the elements in the matrix or more explicitly
\begin{align}
\label{eq:totalcounts}
n=\sum_{i}\sum_{j}M_{ij}.
\end{align}
In this case, the total number of events $n$ may be related to the integrated volume under the fit function to the 2D histogram by a relationship to the normalization factor. This relationship may be easily derived for a 1D distribution and transferred to 2D.

\subsubsection{Derivation using 1D Gaussian}
For any order distribution, $n$ is the total number of counts in the histogram. In the 1D histogram, the entire area of all of the histogram bars added together corresponds to the area underneath the fitted Gaussian distribution. Recall the quantity $\Delta x$ is the bin width of the histogram and has dimensions equivalent to the dimensions of the random variable $x$ and determines the resolution of the histogram. For our purposes both $x$ and $\Delta x$ have spacial dimensions of length.
The form for a Gaussian fit of a 1D histogram is given by
\begin{align}
\label{eq:basicgaussian}
F(x) = A e^{-\frac{(x-\mu)^2}{2\sigma^2}},
\end{align}
where $A$ is the amplitude of the fit, $\mu$ is the peak of the fit, and $\sigma$ is the standard deviation of the distribution. Given that this is a least squares fit to the above histogram, we may quite clearly conclude by looking at Figure~\ref{fig:gaussfit1d} that the area of the histogram bars is just about equal to the area under the fit function. Mathematically, we may express this relationship between the area of the histogram bars and the area under the Gaussian fit as
\begin{align}
\label{eq:sumformofgaussianint}
\sum_i M_i \Delta x \simeq\int_x F(x) \,dx.
\end{align}
as $n$ increases and as $\Delta x$ goes to zero, this statement gets more true---the lefthand side becomes a more accurate prediction of the righthand side. This is basically a statement of the central limit theorem applied to a Riemann sum approximation~\cite{riemann1868}. Upon substituting the 1D analog of (\ref{eq:totalcounts}) into the lefthand side of the above equation to eliminate the sum notation and dividing both sides by $n\Delta x$, we find a direct relationship between the 1D fitted and normal distributions that share the same $\mu$ and $\sigma$, which may be expressed as
\begin{align}
\label{eq:fittednormalrelationfunctions}
N(x)\simeq\frac{F(x)}{n\Delta x}.
\end{align}
This relationship serves as a statistical way to normalize a Gaussian distribution. We get another relationship between the amplitudes of the fitted and normal distributions after substituting the amplitude of the normal distribution $1/\sigma\sqrt{2\pi}$ for $N(x)$ and the amplitude of the fitted distribution $A$ for $F(x)$. Moving all of the factors onto one side yields the following equation containing all the characteristics of the histogram distribution:
\begin{align}
\label{eq:eq_approx}
\frac{A\sigma \sqrt{2\pi}}{n\Delta x}\simeq1.
\end{align}
This statement becomes more accurate in the continuous limit where $n\rightarrow\infty$ and $\Delta x\rightarrow 0$. Note also that since we are binning the $n$ events into a histogram as $n$ approaches infinity the amplitude $A$ of the fit function also approaches infinity, however, concurrently decreasing $\Delta x$ reduces the amplitude of the fit equivalently. As $n\rightarrow\infty$ and $\Delta x$ stays constant we also have $A\rightarrow\infty$. Arguably, since $n$ is the total number of events binned into the entire histogram we will also always have the condition $A<n$. From the point of view of dimensional analysis the dimensions of $A$ are the same as that of $n$ or $[A]=[n]$. We also have from Gaussian statistics that $[\sigma]=[\Delta x]$, so the left hand side of (\ref{eq:eq_approx}) is dimensionless.

\subsubsection{Bivariate distribution}
For the 2D case, we must define two bin widths $\Delta x$ and $\Delta y$ to be the bin width of either dimension of the 2D histogram or random variables $x$ and $y$. For the remainder of this paper we shall assume a symmetrical histogram, that is $\Delta y=\Delta x$ such that lateral cross-sectional area of each bin is $\Delta x^2$.

For the bivariate Gaussian distribution we may write down the 2D analog of (\ref{eq:sumformofgaussianint}) as
\begin{align}
\sum_i\sum_j M_{ij}\Delta x^2 \simeq \int_x\int_y F(x,y)\,dy\,dx,
\end{align}
which may be manipulated to obtain the 2D analog of (\ref{eq:fittednormalrelationfunctions}) to be
\begin{align}
\label{eq:generaldiscretenormalrelation}
N(x,y)\simeq\frac{F(x,y)}{n\Delta x^2}.
\end{align}
This relationship will be used in a crucial step for a proof carried out later in this report.

\section{Direction Algorithm}
\label{sec:directionalgorithm}
To describe the problem at hand, we are faced with a matrix $M$, say for example the $3\times 3$ matrix
\begin{align}
M
=
\begin{pmatrix}
  M_{11} & M_{12} & M_{13} \\
  M_{21} & M_{22} & M_{23} \\
  M_{31} & M_{32} & M_{33} \\
\end{pmatrix}
=
\begin{pmatrix}
  1 & 4 & 2 \\
  0 & 8 & 4 \\
  1 & 0 & 1 \\ 
\end{pmatrix}
\implies 
\begin{pmatrix}
  1 & 4 & 2 \\
  0 & \nearrow & 4 \\
  1 & 0 & 1 \\ 
\end{pmatrix}.
\end{align}
We immediately notice that this matrix has some inherent directionality. Given that the origin of the overlying coordinate system is placed at the center of this matrix (i.e. the center of $M_{22}$) we can tell that it is biased toward the upper right corner elements (i.e. $M_{12}$, $M_{13}$, and $M_{23}$)---we can clearly see the direction of the bias in the matrix $M$. The ultimate goal of the algorithm described here is to find this direction. If we were to plot this matrix as a 2D histogram and fit it with a bivariate Gaussian distribution, the centroid relative to the center of the matrix holds directional information about the matrix. This is a key point for moving forward here. In mathematical terms, since we define a center of the matrix, the location of the centroid of the fitted distribution may be equivalently represented in polar coordinates. The directionality is primarily determined by the angle $\vartheta$ at which the centroid of a 2D Gaussian fit to $M$ is located. The goal thus becomes to find this angle. Essentially, the way the algorithm works is that, given a set of criteria defined by the physical system that generated the matrix $M$, we compare a matrix with known direction, called the reference dataset say $M_\vartheta$, with $M$ using the FND $\|M-M_\vartheta\|_F$, rotating the simulated data about the defined origin to find the closest match with $M$. The reconstructed direction is the angle for which the FND is at its minimum.

The direction algorithm presented here is used to determine the direction in a matrix such as $M$ and consists of the following steps:
\begin{enumerate}
\item Determine the physics of the phenomenon that produced the matrix $M$. Create a simulation based on this model that produces a raw set of $x$ and $y$ coordinates that will constitute or approximate the distribution in $M$. This step may also be replaced by making a 2D measurement under known conditions (i.e. known source direction) with rotatable coordinates. This step produces the reference dataset.
\item Then the reference $x$ and $y$ data points are rotated from $\vartheta\in[0,2\pi]$, where a new dataset of points is produced by each rotation about the circle (e.g. $\vartheta=1^\circ,2^\circ,\dots,360^\circ$) and is binned into a 2D histogram to produce a matrix $M_{1^\circ},M_{2^\circ},\dots,M_{360^\circ}$.
\item For each rotation about the full circle of $2\pi$ radians the Frobenius norm of the difference is calculated between the measured data $M$ and the reference data $M_\vartheta$. These FND calculations are saved and plotted for all the angles in $2\pi$. Ideally, this produces a single plot that has a clear minimum. 

\item The final step is to determine the minimum of this plot. This entails performing a fit of the FND data. Analytically, we may derive a continuous function that models the FND output. This is derived below in Section~\ref{subsec:cfndintro}. The minimum of this fit is the reference angle $\vartheta=\vartheta_0$. As shown in Section~\ref{subsubsec:firstorderapprox} in the first-order limit, the fit of the FND between two similar Gaussian distributions is an absolute sine function, otherwise for higher-order approximations the full analytical form may be used.
\end{enumerate}
The FND is used here as the primary tool to quantify matrix similarity. The general idea behind the algorithm is that at the reference angle, the FND between the measured dataset and the reference dataset will reach a minimum value becasue those datasets are identical. The key revolution about this algorithm is that it takes into account extra information about the system that is simulated and compared to real data, rather than merely performing a single fit on the raw histogram dataset. If the matrix is all the information used to extract the direction, then it is sufficient to perform a least squares fit to the data to find the centroid $\bm \mu$ and thus the direction $\vartheta$ calculated from the centroid. However, by comparing the dataset to binned rotatable coordinates we can potentially extract more directional information.

This algorithm takes advantage of the information that is recorded in the raw data measurements of a physical system and compares it to simulated data. Rather than performing analysis directly on the matrix $M$, this algorithm simulates the physics that produces $M$ to produce continuous coordinates of events that are rotated at angle of the whole circle $2\pi$ and compared with $M$, using the minimum of the FND between measured and simulated datasets to find the direction. This has potential to perform better than a pure fit because we are using more information (the continuous $x$ and $y$ coordinates of raw simulated data) rather than fitting a Gaussian to a single histogram. The drawback of this method is that  the quality of the result requires that either we must understand the physics of what is going on very well or we have a good experiment with a known source.

Another point to mention is that by taking FND comparisons at many angles averages out potential uncertainties in the reconstructed angle. We are in a sense averaging all of the imperfections in the discretization of the data by making many comparisons around the circle and fitting them with an analytical curve. The analytical form of the CFND tells us that the reconstructed angle should be located at the minimum of the absolute sine function.

\subsection{Continuous Frobenius norm of the difference}
\label{subsec:cfndintro}
The FND quantifies the discrepancy between two matrices. In the algorithm described above we treat matrices as 2D histograms of our datasets on which we apply the FND. As described above in Section~\ref{sec:directionalgorithm}, the direction algorithm utilizes the fit of the FND to determine the direction of the matrix. We may define $M_{\vartheta_0}$ to be this matrix dataset at an unknown angle $\vartheta_0$ and $M_\vartheta$ to be the matrix produced by rotating the raw simulated data from $\vartheta_0$ by an angle $\vartheta$ and then binning the rotated events as described in the direction algorithm. Thus, as a first step in the algorithm, we may produce discrete FND data as a function of $\vartheta$ by applying the Frobenius norm in (\ref{eq:frobeniusnorm}) to the difference between these two matrices. We shall define the normalized matrix by dividing the matrix $M_\vartheta$ by the total number of event stored in the matrix $n$, or more explicitly
\begin{align}
\label{eq:normalizedmatrix}
m_{\vartheta}=\frac{M_\vartheta}{n},
\end{align} 
For the normalized unknown dataset $m_{\vartheta_0}$ and normalized simulated dataset $m_\vartheta$, the FND is expressed as
\begin{align}
\label{eq:frobeniusnormdiffform}
\text{FND}\equiv\|M_{\vartheta_0}-M_\vartheta\|_F=\sqrt{\sum_{i}\sum_{j}(m_{\vartheta_0,ij}-m_{\vartheta,ij})^2},
\end{align}
which quantifies how similar the matrices are\footnote{The FND closely resembles the form of the $\chi^2$ test, which for an observed set of numbers $x_i$ and an expected set of numbers $m_i$ we have the $\chi^2$ distribution $$\chi^2=\sum_{i=1}^{k}\frac{(x_i-m_i)^2}{m_i},$$ which can produce plots similar to the square of the FND in (\ref{eq:frobeniusnormdiffform}) if applied correctly.}. The lower the value of the function, the higher the similarity. Thus the minimum of this function, particularly the minimum of the fit to these FND data points, is the closest match angle. Thus, if we produce $M_\vartheta$ for many angles in $2\pi$, we may use the FND data to find the location of the minimum of the FND $\|M_{\vartheta_0}-M_\vartheta\|_F$ to find the unknown angle $\vartheta_0$.

The CFND is defined as the continuous analog of the FND and is designed to do the same as the FND, swapping discrete matrices with continuous 2D distributions. To appropriately define the CFND, we might make some assumptions. We shall suppose that the normalized distribution is closely modeled by a bivariate normal distribution  as $\Delta x \rightarrow 0$ and $n\rightarrow\infty$, whose centroid is located some radial distance $\mu$ from the origin of rotation (center of the matrix). That is the elements of the matrix $m_{\vartheta}$ are like samples of a normal distribution as the resolution of the histogram increases. This bivariate normal distribution we shall call $N_\vartheta(\bm r)$ is defined as a symmetrical bivariate normal distribution as in (\ref{eq:bivariatenormaldistribution}) that has a centroid that is rotated about the origin by an angle $\vartheta$. We simply convert the centroid to polar coordinates to be expressed as
\begin{align}
\bm \mu=\begin{pmatrix}
\mu \cos{\vartheta} \\
\mu \sin{\vartheta}
\end{pmatrix},
\end{align}
where $\mu$ is the distance from the centroid to the origin of rotation. Like this, the continuous Frobenius norm of the difference or CFND expressed as the continuous analog of the FND in (\ref{eq:frobeniusnormdiffform}) and may thus be defined as
\begin{align}
\label{eq:definitionofcfnd}
\text{CFND} \equiv \left[\int_{y_1=-\infty}^{y_2=\infty}\int_{x_1=-\infty}^{x_2=\infty}(N_{\vartheta_0} - N_\vartheta)^2\,dx\,dy\right]^\frac{1}{2},
\end{align}
which when evaluating this integral over all space we obtain an analytical expression that evaluates to a relatively simple function in terms of $\vartheta$. The bounds $x_1,x_2$ and $y_1,y_2$ are included in the integral because if one is analyzing a region that does not include the entire distribution, but is confined to these bounds, one may compare the CFND over the same spatial region numerically. However, the CFND integral must be evaluated over all space in order to conveniently express the integral as an analytical function. The integral over all space, however, assumes that the distribution is entirely contained in the 2D integrating space. This is a reasonable assumption as in many practical applications all of the information contained in the distributions can generally be measured. The CFND integral evaluates to the following analytical expression:
\begin{align}
\label{eq:analyticalexpression}
\text{CFND}
=
\frac{1}{\sigma\sqrt{2\pi}}\left[1-\exp{\left(-\frac{\|\bm \mu_0 - \bm \mu\|^2}{4 \sigma^2}\right)}\right]^\frac{1}{2},
\end{align}
whose derivation is briefly discussed in Appendix~\ref{analyticalevaluation}. Upon expanding $\|\bm \mu_0 - \bm \mu\|^2$ we find that the analytical expression for the CFND simplifies to
\begin{align}
\label{eq:cfndanalyticalformula}
\text{CFND}=\frac{1}{\sigma\sqrt{2\pi}}\left[
1-\exp{\left(\frac{\mu^2(\cos{(\vartheta_0-\vartheta)}-1)}{2\sigma^2}\right)}\right]^\frac{1}{2}.
\end{align}
When $\vartheta_0=\vartheta$, this function equals zero. This function resembles a kind of inverted Laplace distribution, and its relationship to the Laplace distribution requires further study. Using the above equation, in practical applications this allows us to find the unknown direction $\vartheta_0$ by locating the minimum of an analytical fit to the discrete FND dataset that has the form of (\ref{eq:cfndanalyticalformula}).

\begin{figure}[h!]
\begin{tikzpicture}[scale=1]

\node at (3.5,1.5) {\includegraphics[width=1.2in]{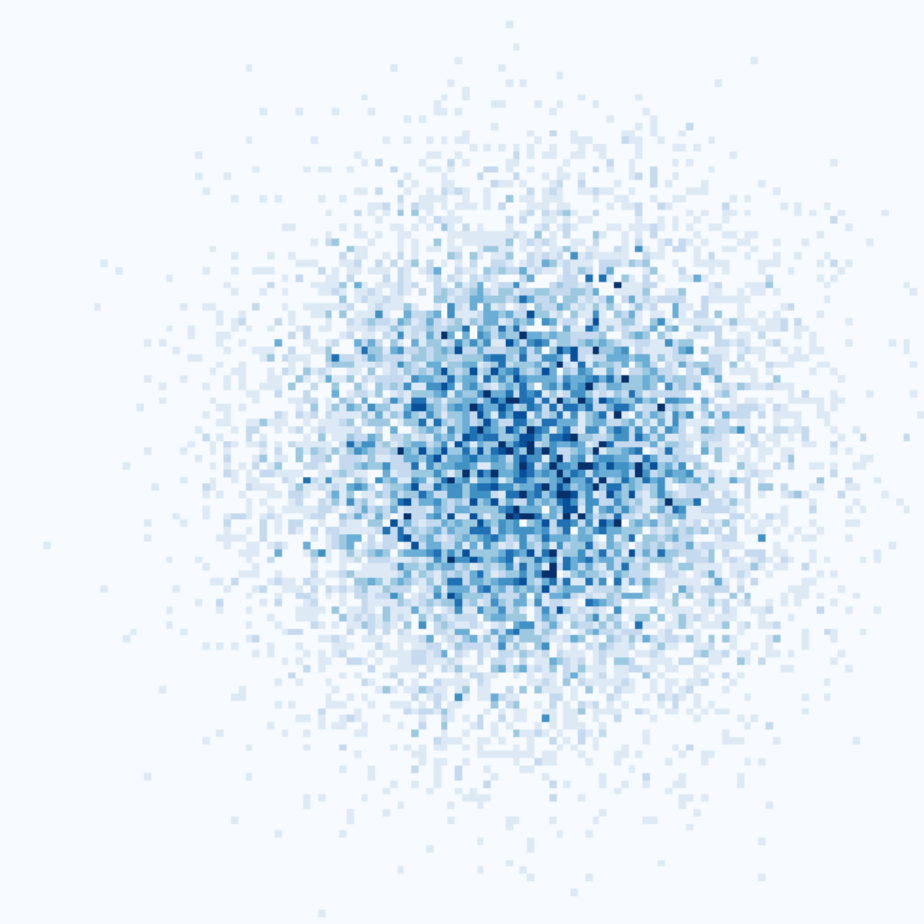}};

\node at (3.5+4.5,1.5) {\includegraphics[width=1.2in]{figs/gauss_data_theta0.pdf}};

\node at (3.5,-2.5) {\includegraphics[width=1.2in]{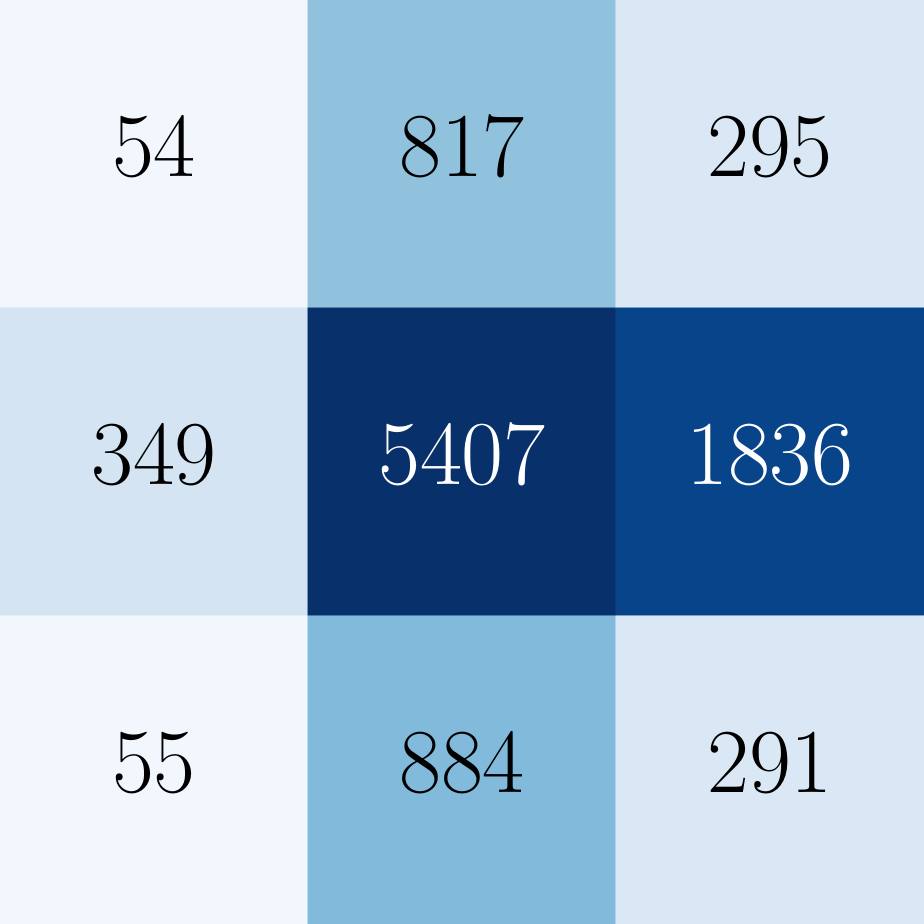}};

\node at (3.5+4.5,-2.5) {\includegraphics[width=1.2in]{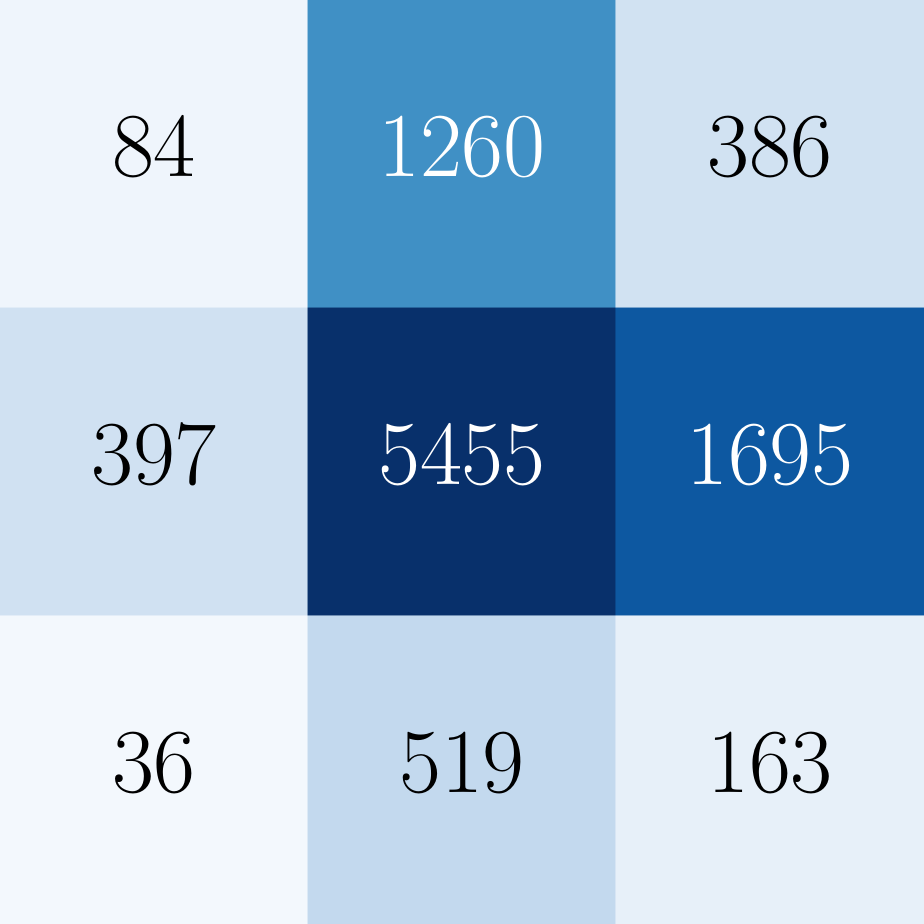}};

\begin{scope}[shift={(2,0)}]
% Original grid
\foreach \x in {0,1,2,3} {
    \draw[gray] (\x,0) -- (\x,3);
}
\foreach \y in {0,1,2,3} {
    \draw[gray] (0,\y) -- (3,\y);
}

\begin{scope}[shift={(1.5,-2.5)}, rotate=0, shift={(-1.5,-1.5)}]
            \foreach \x in {0,1,2,3} {
                \draw[gray] (\x,0) -- (\x,3);
            }
            \foreach \y in {0,1,2,3} {
                \draw[gray] (0,\y) -- (3,\y);
            }
        \end{scope}

\begin{scope}[shift={(1.5+4.5,-2.5)}, rotate=0, shift={(-1.5,-1.5)}]
            \foreach \x in {0,1,2,3} {
                \draw[gray] (\x,0) -- (\x,3);
            }
            \foreach \y in {0,1,2,3} {
                \draw[gray] (0,\y) -- (3,\y);
            }
        \end{scope}

\begin{scope}[shift={(1.5+4.5,1.5)}, rotate=0, shift={(-1.5,-1.5)}]
            \foreach \x in {0,3} {
                \draw[gray!30!white] (\x,0) -- (\x,3);
            }
            \foreach \y in {0,3} {
                \draw[gray!30!white] (0,\y) -- (3,\y);
            }
        \end{scope}

\begin{scope}[shift={(1.5,1.5)}]
% Axes
\draw[->, thick] (0,0) -- (2,0) node[right] {$x$};
\draw[->, thick] (0,0) -- (0,2) node[above] {$y$};
\end{scope}
\end{scope}

% Rotated grid about its center (1.5, 1.5)
\begin{scope}[shift={(1,0)}]
    \begin{scope}[shift={(5.5,0)}] % Move the whole rotated system to the right
        \begin{scope}[shift={(1.5,1.5)}, rotate=-30, shift={(-1.5,-1.5)}]
            \foreach \x in {0,1,2,3} {
                \draw[gray] (\x,0) -- (\x,3);
            }
            \foreach \y in {0,1,2,3} {
                \draw[gray] (0,\y) -- (3,\y);
            }
        \end{scope}
        
        % Rotated axes
        \draw[->, thick, black] (1.5,1.5) -- ++({2*cos(-30)},{2*sin(-30)}) node[right] {$x'$};
        \draw[->, thick, black] (1.5,1.5) -- ++({-2*sin(-30)},{2*cos(-30)}) node[above right] {$y'$};
    \end{scope}
\end{scope}

\draw [decorate,decoration={brace,amplitude=4pt,raise=2pt}]
    (2,0) -- (2,1) node [midway, left=4pt] {\footnotesize $\Delta x$};

\draw [decorate,decoration={brace,amplitude=4pt,raise=2pt}]
    (3,0) -- (2,0) node [midway, below=4pt] {\footnotesize $\Delta x$};

\node [] at (3.5+0.21,1.5) {\textcolor{red}{$\times$}};

\node [] at (3.5+0.21+4.5,1.5) {\textcolor{red}{$\times$}};

\end{tikzpicture}
\caption{Shows affine rotation of 2D histogram bin grid rotation by $\vartheta=30^\circ$ in the $3\times3$ case. The Gaussian data is plotted in the top row with the respective coordinate system and the binned result is shown in the bottom row. This figure highlights the 2D histogram binning process after rotation of the dataset. The centroid of the distribution is marked by a red crosshair.}
\label{fig:nsimrotgauss}
\end{figure}

\subsubsection{First-order approximation}
\label{subsubsec:firstorderapprox}
As an example of using the CFND in a mathematical context, here we focus on a Gaussian distribution for which we are dealing with a very small $\mu$ relative to $\sigma$. In other words, this is the case for which $\mu \ll \sigma$ or equivalently, in regards to (\ref{eq:analyticalexpression}), the coefficient $\mu^2/2\sigma^2 \ll 1$. If this is the case for the distribution dataset, then it is possible to simplify the result by applying the first-order Taylor series approximation for the exponential ($e^\lambda=1+\lambda+\cdots$) to the exponential quantity in (\ref{eq:analyticalexpression}). Upon applying the Taylor series  approximation, for the exponential in (\ref{eq:analyticalexpression}) alone we receive
\begin{align}
\exp{\left(\frac{\mu^2(\cos{(\vartheta_0-\vartheta)-1})}{2\pi\sigma^2}\right)}
\simeq
1+\frac{\mu^2(\cos{(\vartheta_0-\vartheta)-1})}{2\pi\sigma^2}.
\end{align}
Now, inserting this into the analytical CFND expression in (\ref{eq:analyticalexpression}) as well as the half-angle trigonometric identity for the sine function that arises in the equation, we obtain an absolute sine function for the CFND upon taking the square root. In its final form, the first-order approximation of the CFND is written as
\begin{align}
\label{eq:abssineapprox}
\text{CFND}
\simeq
\frac{\mu}{\sqrt{2\pi}\,\sigma^2}\left|\,\sin{\left(\frac{\vartheta_0-\vartheta}{2}\right)}\right|.
\end{align}

\subsubsection{First-order Cauchy approximation}
There are several assumptions that come with this result. First, we assume that the data is closely modeled by Gaussian distributions, which we choose as they appear widely in modern physics. We also suppose that the distributions have no correlation between $x$ and $y$. The CFND integral may be too difficult to analytically evaluate if this was not the case or $\rho\neq 0$.

The CFND is a mathematical construct that may be applied to any distribution or 2D function. There is even the possibility of a higher-dimensional CFND, but this is not discussed in this paper and requires further study. Theoretically, we did not have to choose a Gaussian distribution. In fact, we expect any bell-shaped distribution to lead to an absolute sine function in the first-order approximation. Perhaps, even a wider variety of functions will produce sinusoidal or other simple functions. Here the CFND is calculated in the same way as in the previous section but modeling the data using Cauchy distributions, or Lorentz distributions. The integral of the radially isotropic bivariate Cauchy distribution is analytically too difficult to solve because the expansion of the product of two bivariate Cauchy distributions is too high order---we do not yet know the analytical solution. Rather, we may use a separated bivariate Cauchy distribution in the CFND integral, which has the form
\begin{align}
N_\vartheta(x,y)
=
\frac{1}{\pi^2\gamma^2}
\frac{1}{
\left[1+\left(\frac{x-\mu_x}{\gamma}\right)^2\right]
\left[1+\left(\frac{y-\mu_y}{\gamma}\right)^2\right]
},
\end{align}
where $\gamma$ is a measure of the width or spread of the Cauchy distribution and has the same dimensions of $x$. When appropriately integrating using (\ref{eq:definitionofcfnd}) to find the CFND we get an elaborate analytical result. This result is typeset and briefly discussed in Appendix~\ref{sec:cauchyderivation}. Remarkably, when taking the first-order approximation where $\mu\ll\gamma$ as we did in the Gaussian case, all but one term is eliminated which may be simplified to the absolute sine function
\begin{align}
\label{eq:cfndcauchyapprox}
\text{CFND}_\text{Cauchy}\simeq
\frac{\mu}{\sqrt{2}\pi\gamma^2}\left|\,\sin{\left(\frac{\vartheta_0-\vartheta}{2}\right)}\right|.
\end{align}
We have shown the first-order approximation approaches an absolute sine function for both the normal distribution and the Cauchy distribution. We propose that any distribution closely modeled by a Gaussian distribution will yield an absolute sine function in the first-order approximation. The shape of both the Gaussian and Cauchy distributions are similar and both simplify to an absolute sine function in the first-order approximation. We are limited by the inability of the mathematics notation---specifically the calculus integral---to handle the complexity of other functions in the CFND.

\subsection{Relating CFND and FND}
The CFND must be related to the FND in order to apply it in the direction algorithm. The CFND was created by applying an idea of continuity to the FND---the discretely normalized histogram distributions were replaced with continuous normal distributions corresponding to the Gaussian fits of the matrices, and a housekeeping modification was made by replacing the sums over the indices of the matrices in the FND with integrals over the corresponding spacial dimensions of the data. However, although the FND and CFND have a similar form we may derive their precise correlation using the fitted-normal relationship derived in (\ref{eq:generaldiscretenormalrelation}) and dimensional analysis.

The normal distribution in 2D has dimensions of $1/L^2$. Thus, due to the double integral in the CFND, in 2D the CFND has dimensions of $1/L$. The FND has units of counts and is dimensionless since it is just the addition of histograms. The approach to relating the CFND and the FND is simple. We shall simply insert the expression of $M$ in (\ref{eq:generaldiscretenormalrelation}) into (\ref{eq:frobeniusnormdiffform}) to recieve the approximate form of the CFND. For this we require that for the total number of events $n$ binned in the matrix $M$, we have the normalized matrix $m_\vartheta$ as in (\ref{eq:normalizedmatrix}), where $n$ is the total number of events binned into the matrix histogram $M_\vartheta$. Inserting the normalized matrices into (\ref{eq:frobeniusnormdiffform}), adding units to the sum to mirror the integrals of the CFND and to preserve dimensions, and finally substituting (\ref{eq:generaldiscretenormalrelation}) in for the matrices as shown in the above equation, we receive the following relationship:
\begin{align}
\text{CFND} \simeq \left[\sum_{i}\sum_{j}\left(\frac{m_{\vartheta_0,ij}}{\Delta x^2}-\frac{m_{\vartheta,ij}}{\Delta x^2}\right)^2 \Delta x^2\right]^\frac{1}{2},
\end{align}
where $m_{\vartheta_0}$ and $m_{\vartheta}$ are the matrix normalized reference and simulated datasets, respectively. Upon simplifying the above equation and noticing that the FND in (\ref{eq:frobeniusnormdiffform}) may be factored out, we obtain the following relationship between the CFND and the FND:
\begin{align}
\label{eq:cfndfndrelation}
\text{FND}\simeq \Delta x\cdot \text{CFND}.
\end{align}
This formula relates the discrete sum of the matrix elements of the Frobenius norm in (\ref{eq:frobeniusnormdiffform}) to a continuous integral over the equivalent distribution space (\ref{eq:definitionofcfnd}). So this statement becomes more true as $n$ is large and $\Delta x$ is small, invoking the general concept of the central limit theorem. Note that the CFND in (\ref{eq:definitionofcfnd}) has dimensions of $1/L$, so multiplying the CFND by $\Delta x$ which has units of $L$, both sides of the above equation are dimensionless as the FND is a dimensionless quantity---a mathematical operation on the elements of matrices.

The relationship in (\ref{eq:cfndfndrelation}) is important as it allows us to predict the exact fit of the FND curve produced for a distribution dataset. So by combining the result for Gaussian distributions in (\ref{eq:cfndanalyticalformula}) with the relationship in (\ref{eq:cfndfndrelation}), by essentially preserving units and multiplying (\ref{eq:cfndanalyticalformula}) by $\Delta x$ the predicted fit function becomes
\begin{align}
\text{FND}\simeq\frac{\Delta x}{\sqrt{2\pi}\,\sigma}\left[
1-\exp{\left(\frac{\mu^2(\cos{(\vartheta_0-\vartheta)}-1)}{2\sigma^2}\right)}\right]^\frac{1}{2}.
\end{align}
We may also apply the relationship to the first-order approximation result in (\ref{eq:abssineapprox}) to get the first-order fit function
\begin{align}
\label{eq:finalfitcurve}
\text{FND}
\simeq
\frac{\mu \Delta x}{\sqrt{2\pi}\,\sigma^2}\left|\,\sin{\left(\frac{\vartheta_0-\vartheta}{2}\right)}\right|.
\end{align}
Again, this analytical function will be a better approximation of the FND curve data points as $n\rightarrow\infty$, $\Delta x\rightarrow 0$, and $\mu\ll\sigma$. The above equation is the theoretical fit to our simulated data in these limits, where $\vartheta_0$ is the unknown direction of the measured dataset. In practice the analytical function that best fits the data may require a vertical offset, however, the minimum of the fit function is the reconstructed angle.
\begin{figure}
\centering
\includegraphics[width=0.95\columnwidth]{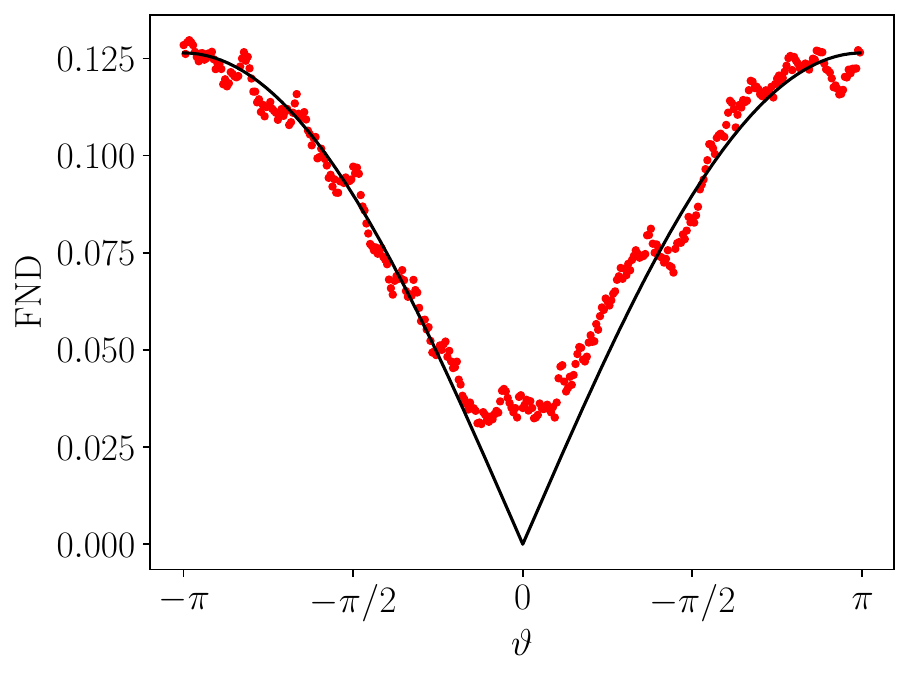}
\caption{\label{fig:fnddatafitfinalresult} Simulation of algorithm on dataset with true or unknown angle $\vartheta_0=0^\circ$. Red is simulated FND data for $n=1000$ and $\Delta x=16$. The distribution parameters were $\sigma=10$ and $\mu=2$. The black curve is the theoretical CFND curve described in (\ref{eq:finalfitcurve}). The minimum of the FND data indicates the reference angle of the dataset.}
\end{figure}

\subsection{Simulation methodology}
\label{subsec:simulations}
Some basic computer calculations were carried out to study the CFND relationship and function behavior. These simulations modelled the described algorithm and entailed producing discrete distributions in Python using the Numpy package for the normal distribution and comparing the FND of the difference between two discrete distributions with the analytical result derived in this paper for different histogram resolutions. The bulk of the simulation results are shown in Figures~\ref{fig:dx16}, \ref{fig:dx8}, \ref{fig:dx4}, and \ref{fig:dx2}. Each of these simulations had a constant resolution defined by the histogram bin width $\Delta x=16,8,4,2$, respectively. The event number $n$ was varied in each simulation for each histogram resolution going up to $n=10^6$ counts. To ensure the integrating space was the same in all simulations the corresponding grid sizes were $128\times128,64\times64,32\times32,16\times16,8\times8$, and $4\times4$. The CFND relationship derived in this report is verified in these simulations of the FND between Gaussian distribution histograms. The FND simulations that are presented there show that for large $n$ and small $\Delta x$ the above theoretical function in (\ref{eq:finalfitcurve}) is an increasingly better fit to the FND data. The bottom right plot in Fig~\ref{fig:simninfty} shows that in the $n\rightarrow\infty$, $\Delta x\rightarrow 0$ limit, the fit function is a perfect fit to the FND data. In other words, the simulation results align well with the theory.

To simulate the case where $n\rightarrow\infty$ we produced a matrix by sampling a 2D Gaussian function to produce the matrix of histogram data varied the bin width $\Delta x$. These simulations are shown in Figure~\ref{fig:simninfty}. Theoretically this is a way to simulate the $n=\infty$ case, in an approximation using a limit. The amplitude of the Gaussian used in this simulation was $A=10$. Note that in this simulation all parameters are constant except for $\Delta x$. This simulation verifies (\ref{eq:analyticalexpression}) and (\ref{eq:finalfitcurve}) in the $n\rightarrow\infty$ and $\Delta x\rightarrow 0$ limit.

For the interest of the reader a set of plots were made of the square of the difference between the reference sampled Gaussian distribution and the distribution rotated by an angle $\vartheta$ in Figure~\ref{fig:nsimdiff}. This shows the magnitude of the difference of two distributions as one is rotated by an angle $\vartheta$. A single point of the FND in the plot in Figure~\ref{fig:fnddatafitfinalresult} is the square root of the sum of all the points of a single histogram in Figure~\ref{fig:nsimdiff}. A maximum FND is achieved at $\vartheta=180^\circ$ and a minimum for the reference angle $\vartheta=\vartheta_0=0^\circ$.

\section{Conclusion}
\label{sec:conclusion}
The theory and method presented in this paper have broad applicability, with a particularly relevant use case in neutrino physics---specifically, in determining the direction of incoming neutrinos using neutron capture locations in segmented detectors. The novel direction algorithm presented in this paper has potential applications in detector neutrino physics, multimodal astronomy, machine learning and more. While this study focuses on rotational comparisons of similar 2D datasets modeled with Gaussian distributions, the framework of the continuous Frobenius norm of the difference (CFND) is general and may be extended to other types of comparison and use different distributions, such as the Cauchy distribution, which was briefly studied in this paper, as well as other physically relevant distributions, models, or functions.

We introduced the CFND as a continuous analog to the discrete Frobenius norm of the difference (FND), derived an analytical form, and demonstrated its use in the algorithm for determining the reference angle $\theta$ of a measured 2D dataset. Our results show that, for bell-shaped curves that may be closely modeled by the Gaussian distribution, the CFND follows a simple absolute sine function behavior, providing a reliable and interpretable metric for estimating directionality in 2D data distributions.

Our simple set of computer simulations highlights a potentially major practical insight. As shown in the figures in Appendix~\ref{sec:simulationresults}, for higher resolution (i.e. smaller $\Delta x$), the FND data converges more slowly to the theoretical fit as $n$ increases. This suggests that experiments with low event rates may be partially offset by larger detector segment sizes, a trade-off for higher angular uncertainty with direct implications for detector design in neutrino experiments. The optimal balance between resolution and event statistics remains an open area for further study. As for other future outlooks, while the work presented in this paper is primarily confined to 2D scalar fields best suited for planar rotations as utilized in the direction algorithm, the CFND framework may easily extend naturally to 3D scalar fields. Although rotation in three dimensions may not always be necessary for directionality extraction, the 3D form of the CFND could prove valuable for other analysis involving volumetric data. Ultimately, the CFND presents a promising tool for quantifying characteristics of discrete data in several scientific domains, from particle physics to data science and beyond.

\section{Acknowledgments}
This work was partially supported by the Consortium for Monitoring, Technology, and Verification. This work was performed under the auspices of the U.S. Department of Energy by Lawrence Livermore National Laboratory under Contract DE-AC52-07NA27344. LLNL-JRNL-2007053.

\subsection{Code availability}
The analysis code used in this work is publicly available at \url{https://github.com/jgyepez/2d-direction-analysis}.

%\bibliography{ref.bib}

\begin{thebibliography}{9}%
\makeatletter
\providecommand \@ifxundefined [1]{%
 \@ifx{#1\undefined}
}%
\providecommand \@ifnum [1]{%
 \ifnum #1\expandafter \@firstoftwo
 \else \expandafter \@secondoftwo
 \fi
}%
\providecommand \@ifx [1]{%
 \ifx #1\expandafter \@firstoftwo
 \else \expandafter \@secondoftwo
 \fi
}%
\providecommand \natexlab [1]{#1}%
\providecommand \enquote  [1]{``#1''}%
\providecommand \bibnamefont  [1]{#1}%
\providecommand \bibfnamefont [1]{#1}%
\providecommand \citenamefont [1]{#1}%
\providecommand \href@noop [0]{\@secondoftwo}%
\providecommand \href [0]{\begingroup \@sanitize@url \@href}%
\providecommand \@href[1]{\@@startlink{#1}\@@href}%
\providecommand \@@href[1]{\endgroup#1\@@endlink}%
\providecommand \@sanitize@url [0]{\catcode `\\12\catcode `\$12\catcode
  `\&12\catcode `\#12\catcode `\^12\catcode `\_12\catcode `\%12\relax}%
\providecommand \@@startlink[1]{}%
\providecommand \@@endlink[0]{}%
\providecommand \url  [0]{\begingroup\@sanitize@url \@url }%
\providecommand \@url [1]{\endgroup\@href {#1}{\urlprefix }}%
\providecommand \urlprefix  [0]{URL }%
\providecommand \Eprint [0]{\href }%
\providecommand \doibase [0]{http://dx.doi.org/}%
\providecommand \selectlanguage [0]{\@gobble}%
\providecommand \bibinfo  [0]{\@secondoftwo}%
\providecommand \bibfield  [0]{\@secondoftwo}%
\providecommand \translation [1]{[#1]}%
\providecommand \BibitemOpen [0]{}%
\providecommand \bibitemStop [0]{}%
\providecommand \bibitemNoStop [0]{.\EOS\space}%
\providecommand \EOS [0]{\spacefactor3000\relax}%
\providecommand \BibitemShut  [1]{\csname bibitem#1\endcsname}%
\let\auto@bib@innerbib\@empty
%</preamble>
\bibitem [{\citenamefont {Duvall}\ \emph {et~al.}(2024)\citenamefont {Duvall},
  \citenamefont {Crow}, \citenamefont {Dornfest}, \citenamefont {Learned},
  \citenamefont {Bergevin}, \citenamefont {Dazeley},\ and\ \citenamefont
  {Li}}]{PhysRevApplied.22.054030}%
  \BibitemOpen
  \bibfield  {author} {\bibinfo {author} {\bibfnamefont {M.~J.}\ \bibnamefont
  {Duvall}}, \bibinfo {author} {\bibfnamefont {B.~C.}\ \bibnamefont {Crow}},
  \bibinfo {author} {\bibfnamefont {M.~A.}\ \bibnamefont {Dornfest}}, \bibinfo
  {author} {\bibfnamefont {J.~G.}\ \bibnamefont {Learned}}, \bibinfo {author}
  {\bibfnamefont {M.~F.}\ \bibnamefont {Bergevin}}, \bibinfo {author}
  {\bibfnamefont {S.~A.}\ \bibnamefont {Dazeley}}, \ and\ \bibinfo {author}
  {\bibfnamefont {V.~A.}\ \bibnamefont {Li}},\ }\href {\doibase
  10.1103/PhysRevApplied.22.054030} {\bibfield  {journal} {\bibinfo  {journal}
  {Phys. Rev. Appl.}\ }\textbf {\bibinfo {volume} {22}},\ \bibinfo {pages}
  {054030} (\bibinfo {year} {2024})}\BibitemShut {NoStop}%
\bibitem [{\citenamefont {Higham}(1988)}]{higham1988matrix}%
  \BibitemOpen
  \bibfield  {author} {\bibinfo {author} {\bibfnamefont {N.~J.}\ \bibnamefont
  {Higham}},\ }\href@noop {} {\emph {\bibinfo {title} {Matrix nearness problems
  and applications}}}\ (\bibinfo  {publisher} {University of Manchester},\
  \bibinfo {year} {1988})\BibitemShut {NoStop}%
\bibitem [{\citenamefont {McElroy}\ and\ \citenamefont
  {Roy}(2021)}]{10.1111/rssb.12480}%
  \BibitemOpen
  \bibfield  {author} {\bibinfo {author} {\bibfnamefont {T.~S.}\ \bibnamefont
  {McElroy}}\ and\ \bibinfo {author} {\bibfnamefont {A.}~\bibnamefont {Roy}},\
  }\href {\doibase 10.1111/rssb.12480} {\bibfield  {journal} {\bibinfo
  {journal} {Journal of the Royal Statistical Society Series B: Statistical
  Methodology}\ }\textbf {\bibinfo {volume} {84}},\ \bibinfo {pages} {473}
  (\bibinfo {year} {2021})}\BibitemShut {NoStop}%
\bibitem [{\citenamefont {Townsend}\ and\ \citenamefont
  {Trefethen}(2015)}]{townsendtrefethen}%
  \BibitemOpen
  \bibfield  {author} {\bibinfo {author} {\bibfnamefont {A.}~\bibnamefont
  {Townsend}}\ and\ \bibinfo {author} {\bibfnamefont {L.~N.}\ \bibnamefont
  {Trefethen}},\ }\href {\doibase 10.1098/rspa.2014.0585} {\bibfield  {journal}
  {\bibinfo  {journal} {Proc. R. Soc. A.}\ } (\bibinfo {year} {2015}),\
  10.1098/rspa.2014.0585}\BibitemShut {NoStop}%
\bibitem [{\citenamefont {Frobenius}(1878)}]{frobenius1878}%
  \BibitemOpen
  \bibfield  {author} {\bibinfo {author} {\bibfnamefont {F.~G.}\ \bibnamefont
  {Frobenius}},\ }\href@noop {} {\bibfield  {journal} {\bibinfo  {journal}
  {Journal für die reine und angewandte Mathematik}\ }\textbf {\bibinfo
  {volume} {84}},\ \bibinfo {pages} {1} (\bibinfo {year} {1878})}\BibitemShut
  {NoStop}%
\bibitem [{\citenamefont {Golub}\ and\ \citenamefont
  {Van~Loan}(1996)}]{golub1996matrix}%
  \BibitemOpen
  \bibfield  {author} {\bibinfo {author} {\bibfnamefont {G.~H.}\ \bibnamefont
  {Golub}}\ and\ \bibinfo {author} {\bibfnamefont {C.~F.}\ \bibnamefont
  {Van~Loan}},\ }\href@noop {} {\emph {\bibinfo {title} {Matrix
  Computations}}},\ \bibinfo {edition} {3rd}\ ed.\ (\bibinfo  {publisher}
  {Johns Hopkins University Press},\ \bibinfo {address} {Baltimore, MD},\
  \bibinfo {year} {1996})\ p.~\bibinfo {pages} {55}\BibitemShut {NoStop}%
\bibitem [{\citenamefont {Gauss}(1809)}]{gaussoriginal}%
  \BibitemOpen
  \bibfield  {author} {\bibinfo {author} {\bibfnamefont {C.~F.}\ \bibnamefont
  {Gauss}},\ }\href@noop {} {\emph {\bibinfo {title} {Theoria motus corporum
  coelestium in sectionibus conicis solem ambientium}}}\ (\bibinfo {year}
  {1809})\ pp.\ \bibinfo {pages} {212--213}\BibitemShut {NoStop}%
\bibitem [{\citenamefont {James}(2006)}]{doi:10.1142/6096}%
  \BibitemOpen
  \bibfield  {author} {\bibinfo {author} {\bibfnamefont {F.}~\bibnamefont
  {James}},\ }\href {\doibase 10.1142/6096} {\emph {\bibinfo {title}
  {Statistical Methods in Experimental Physics}}},\ \bibinfo {edition} {2nd}\
  ed.\ (\bibinfo  {publisher} {World Scientific},\ \bibinfo {year}
  {2006})\BibitemShut {NoStop}%
\bibitem [{\citenamefont {Riemann}(1868)}]{riemann1868}%
  \BibitemOpen
  \bibfield  {author} {\bibinfo {author} {\bibfnamefont {B.}~\bibnamefont
  {Riemann}},\ }in\ \href@noop {} {\emph {\bibinfo {booktitle} {Abhandlungen
  der Königlichen Gesellschaft der Wissenschaften zu Göttingen}}},\
  Vol.~\bibinfo {volume} {13}\ (\bibinfo {year} {1868})\ pp.\ \bibinfo {pages}
  {87--132},\ \bibinfo {note} {section 4 (pp. 101–103) defines the Riemann
  integral via Riemann sums}\BibitemShut {NoStop}%
\end{thebibliography}


\appendix
\section{Gaussian derivation}
\label{analyticalevaluation}
The integral of the CFND using Gaussians in (\ref{eq:definitionofcfnd}) comes down to integrating the square of the difference of two bivariate normalized Gaussian distributions. Analytically, this is done by expanding the square of the difference to get three terms and evaluating the integral of each term separately and summing to get the result. The two squared terms are the same as $N_{\vartheta_0}$ is just a shifted $N_{\vartheta}$ so the integral over all space of $N_{\vartheta_0}^2$ is equal to the integral of $N_{\vartheta}^2$, which is a well known integral that evaluates to
\begin{align}
\int_{-\infty}^\infty\int_{-\infty}^\infty
N_\vartheta^2\,dx\,dy=\frac{1}{4\pi\sigma^2}.
\end{align}
The remaining integral of the cross terms is the integral of the product of two different normal distributions. This integral evaluates to
\begin{align}
\int_{-\infty}^\infty \int_{-\infty}^\infty N_{\vartheta_0}N_\vartheta \, dx \, dy = \frac{1}{4\pi\sigma^2} \exp\left(-\frac{\|\boldsymbol{\mu}_0 - \boldsymbol{\mu}\|^2}{4\sigma^2}\right).
\end{align}
And upon combining these results in the appropriate way that the CFND requires we receive the analytical form that was presented in (\ref{eq:analyticalexpression}).

\section{Cauchy derivation}
\label{sec:cauchyderivation}
The complete analytical expression when evaluating the CFND integral by modeling the data distribution using separated bivariate Cauchy distributions is given by
\begin{widetext}
\small
\begin{align}
\text{CFND}_\text{Cauchy}
=
\frac{1}{\sqrt{2}\pi}
\left[\frac
{
\mu^2\left[
2(16\gamma^2+\mu^2)
+
\mu^2\left(
\cos{(3\vartheta+\vartheta_0)}
+\cos{(\vartheta+3\vartheta_0)}
-2\cos{(\vartheta-\vartheta_0)}
-2\cos{(2(\vartheta+\vartheta_0))}
\right)
\right]\sin^2{\left(\frac{\vartheta-\vartheta_0}{2}\right)}
}
{
\gamma^2
\left[
2(4\gamma^2+\mu^2)
+
\mu^2\left(
\cos{2\vartheta}
-4\cos{\vartheta}\cos{\vartheta_0}
+\cos{2\vartheta_0}
\right)
\right]
[
4\gamma^2+\mu^2\left(
\sin{\vartheta}-\sin{\vartheta_0}
\right)^2
]
}
\right]^\frac{1}{2},
\end{align}
\end{widetext}
which was evaluated using Mathematica analytical integration. We notice, however, that by taking the first-order approximation where $\mu\ll\gamma$ many of the $\mu^2$ terms here are negligible in this approximation and we get the simplified form in (\ref{eq:cfndcauchyapprox}).

\section{Simulation results}
\label{sec:simulationresults}
\begin{figure*}[h]
    \centering
    \subfigure[$\Delta x=32$, $(4\times4)$]{
        \includegraphics[width=0.3\textwidth]{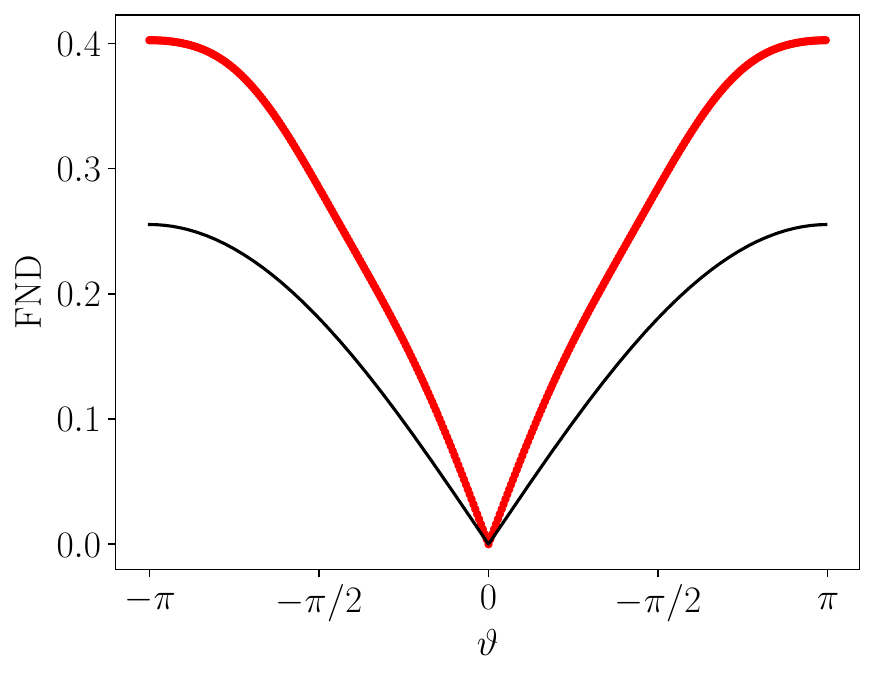}
    }
    \hfill
    \subfigure[$\Delta x=16$, $(8\times8)$]{
        \includegraphics[width=0.3\textwidth]{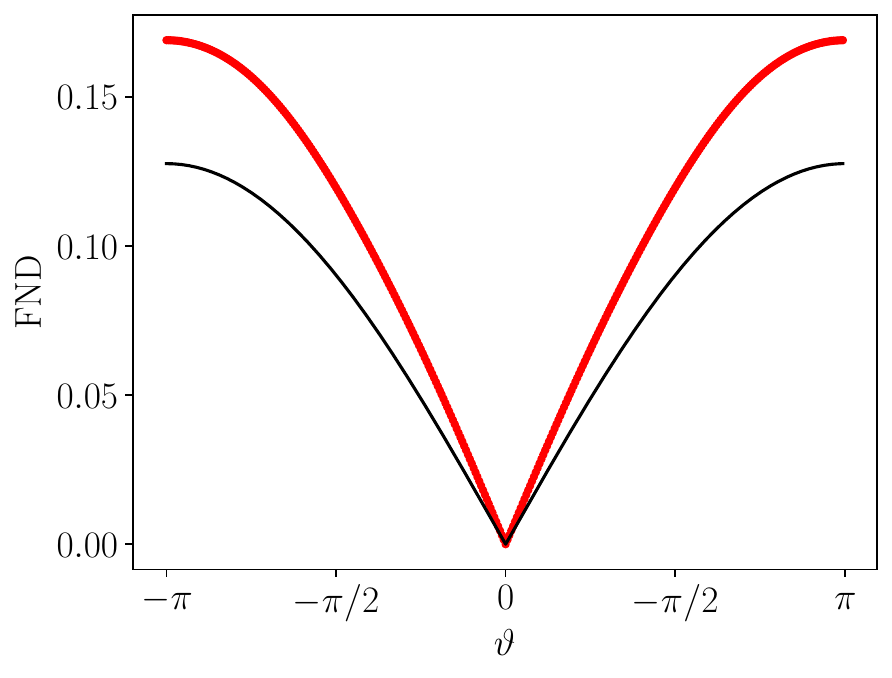}
    }
    \hfill
    \subfigure[$\Delta x=8$, $(16\times16)$]{
        \includegraphics[width=0.3\textwidth]{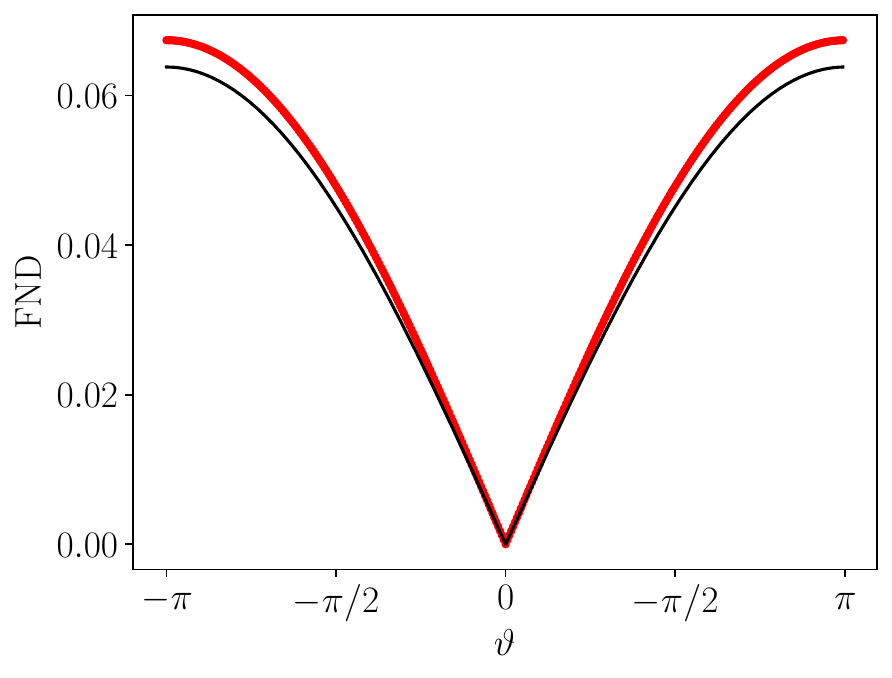}
    }
        \vspace{0.1em}
        \subfigure[$\Delta x=4$, $(32\times32)$]{
        \includegraphics[width=0.3\textwidth]{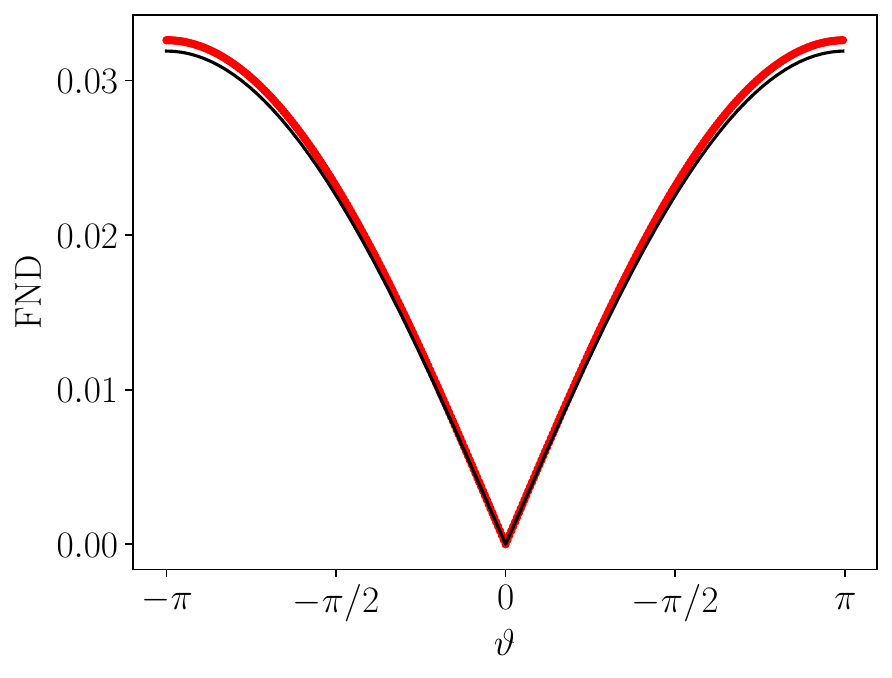}
    }
    \hfill
    \subfigure[$\Delta x=2$, $(64\times64)$]{
        \includegraphics[width=0.3\textwidth]{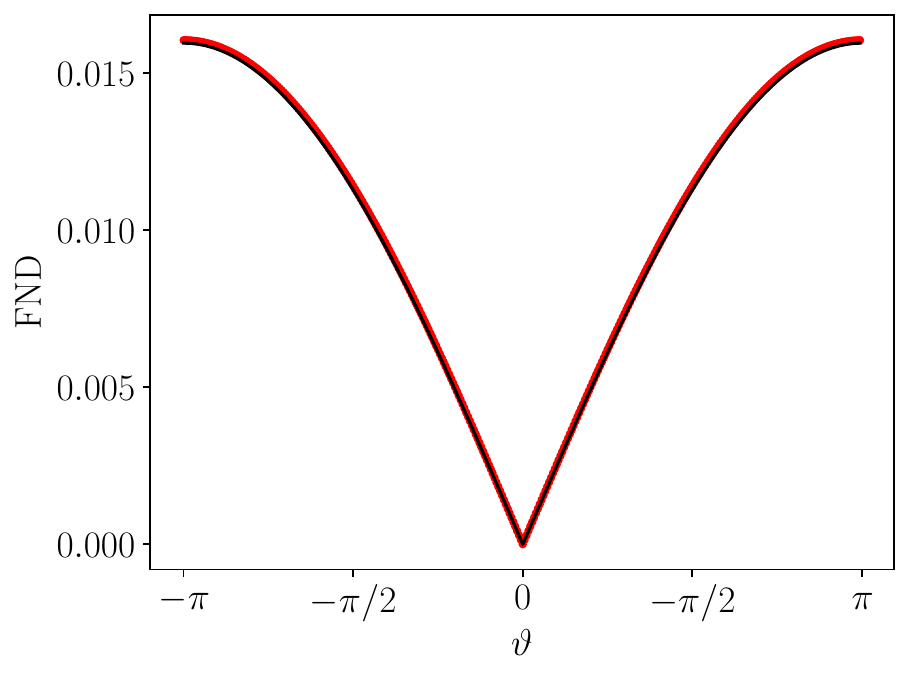}
    }
    \hfill
    \subfigure[$\Delta x=1$, $(128\times128)$]{
        \includegraphics[width=0.3\textwidth]{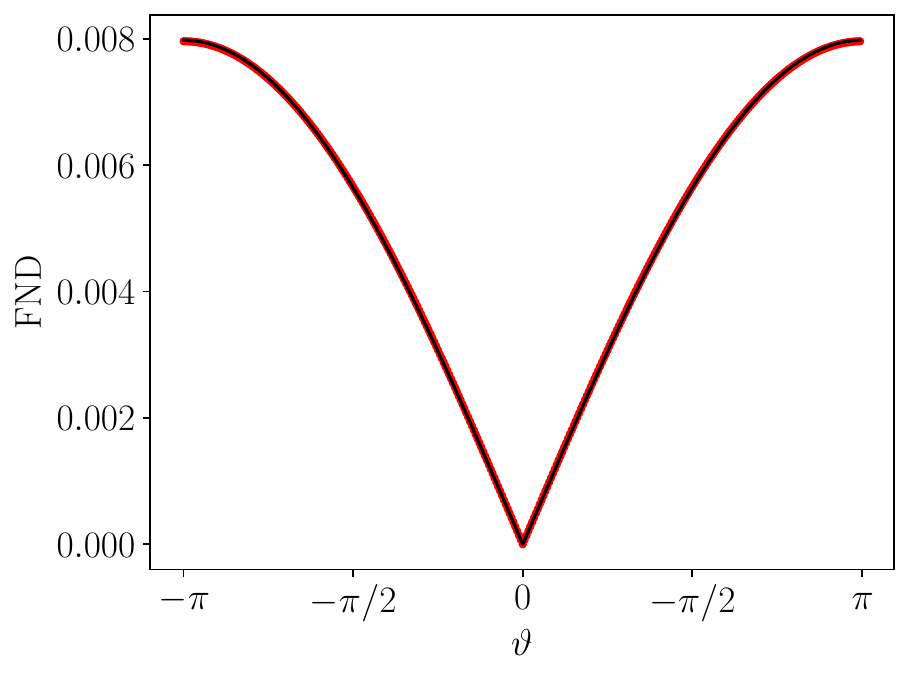}
}
    \caption{Ran six simulations for $\Delta x=1,2,4,6,8,16,32,64$. Corresponding grid sizes were $128,64,32,16,8,4$. Constant parameters were $\sigma=10$ and $\mu=2$. A sampled Gaussian technique was used to model the $n\rightarrow\infty$ case.}
    \label{fig:simninfty}
\end{figure*}

\begin{figure*}[h]
    \centering
    \subfigure[$\vartheta=0^\circ$]{
        \includegraphics[width=0.3\textwidth]{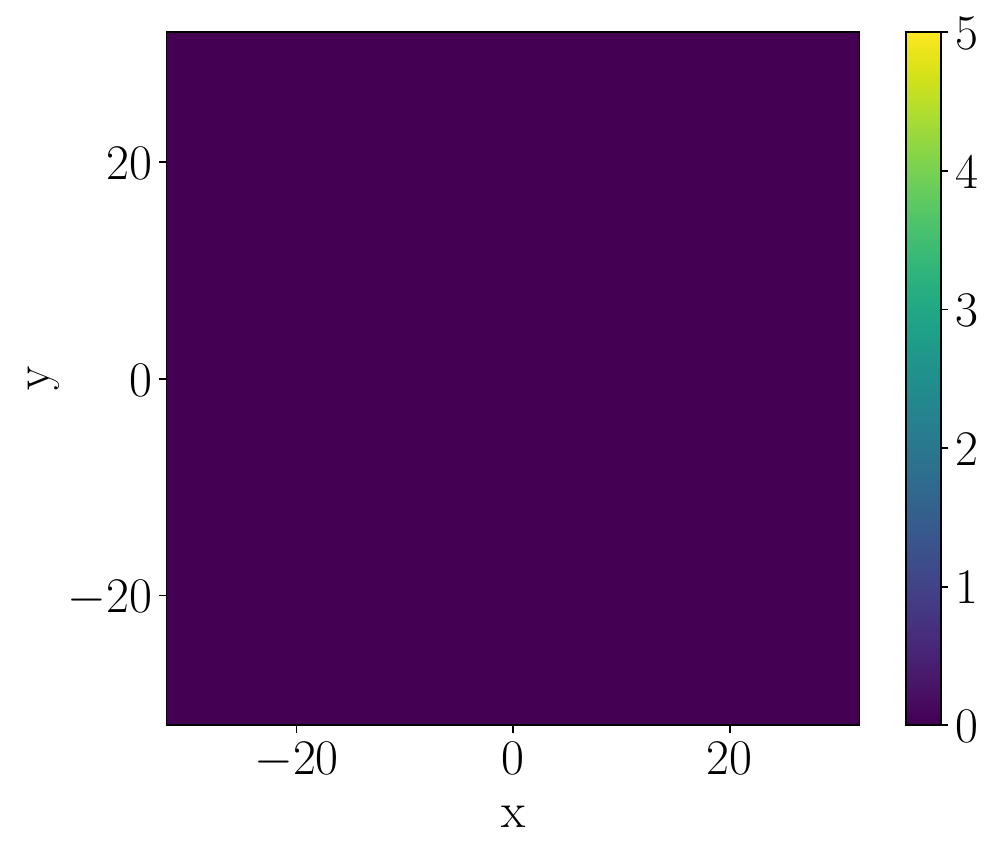}
    }
    \hfill
    \subfigure[$\vartheta=15^\circ$]{
        \includegraphics[width=0.3\textwidth]{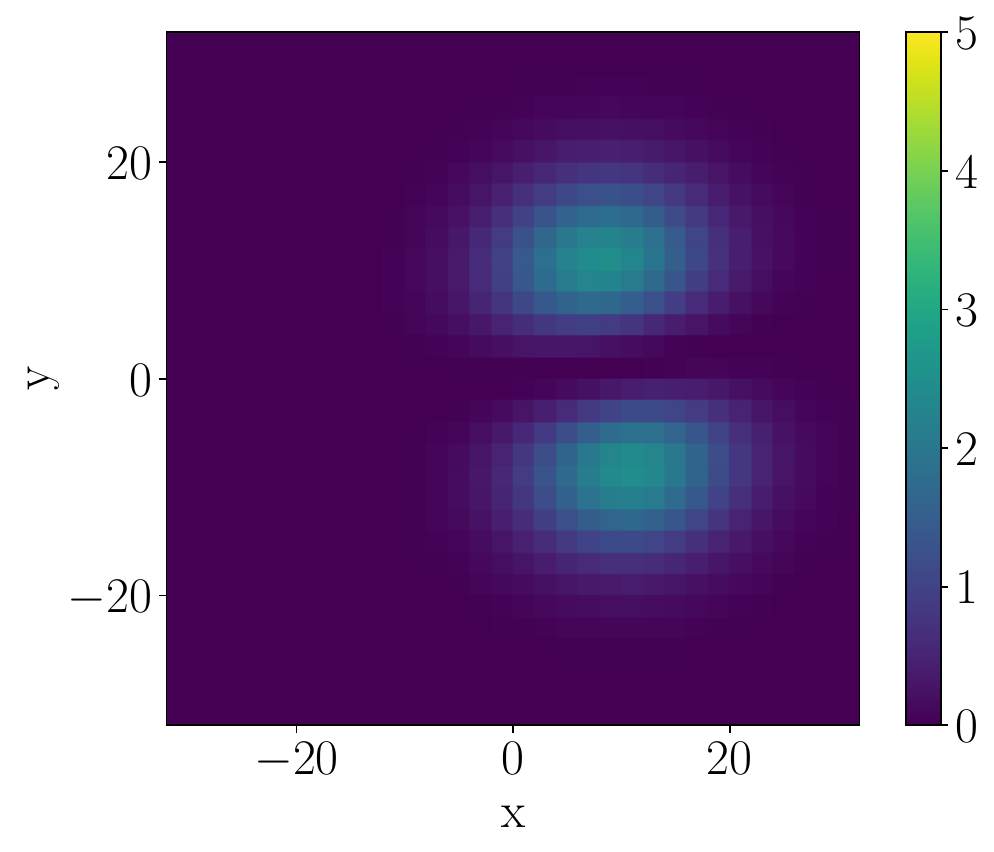}
    }
    \hfill
    \subfigure[$\vartheta=45^\circ$]{
        \includegraphics[width=0.3\textwidth]{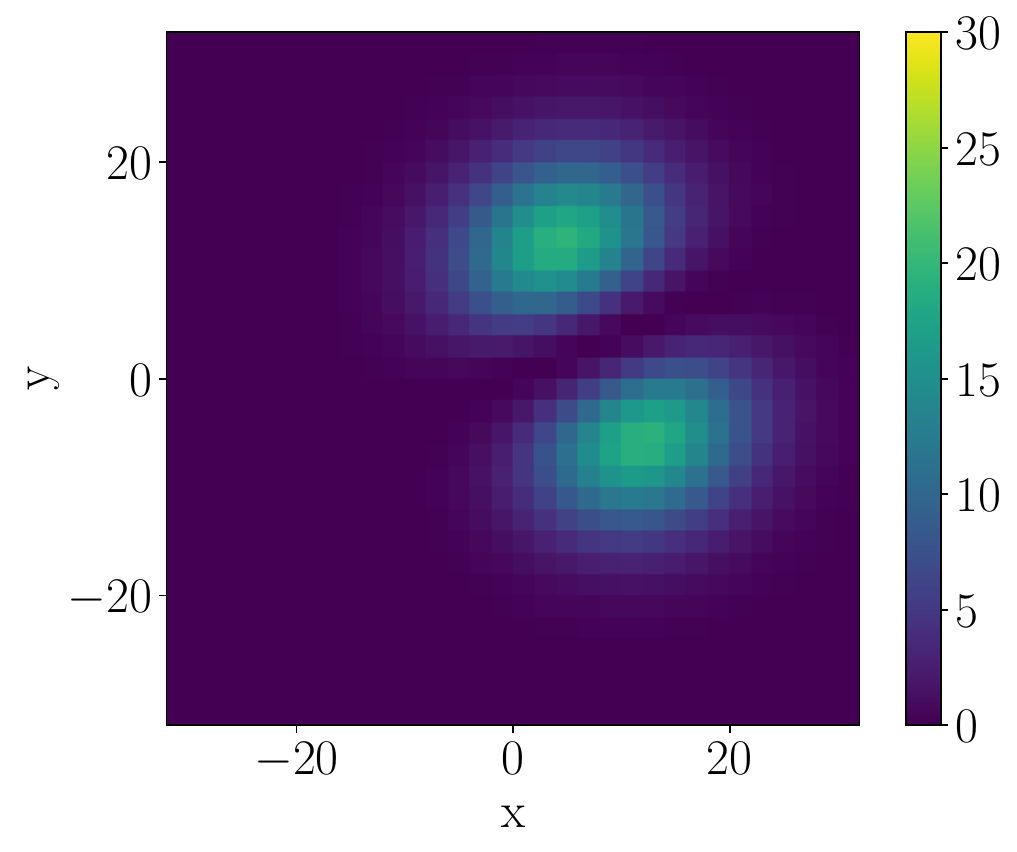}
    }
        \vspace{0.1em}
    \subfigure[$\vartheta=90^\circ$]{
        \includegraphics[width=0.3\textwidth]{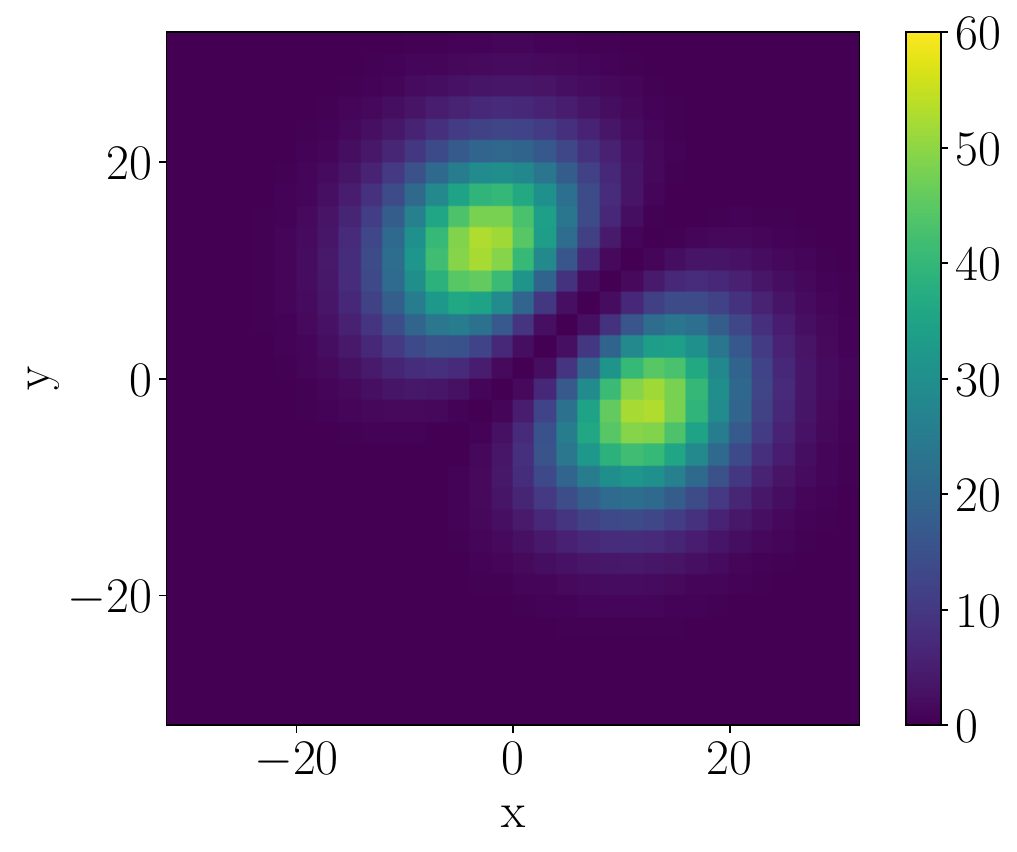}
    }
    \hfill
    \subfigure[$\vartheta=135^\circ$]{
        \includegraphics[width=0.3\textwidth]{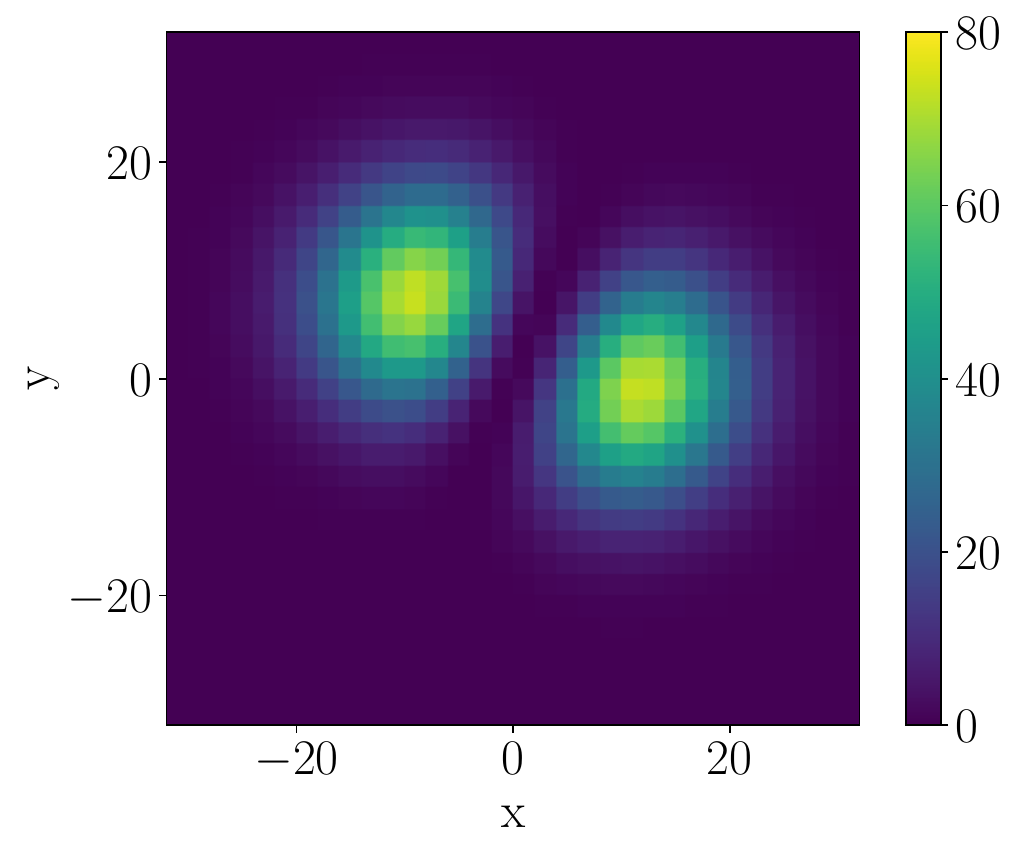}
    }
    \hfill
    \subfigure[$\vartheta=180^\circ$]{
        \includegraphics[width=0.3\textwidth]{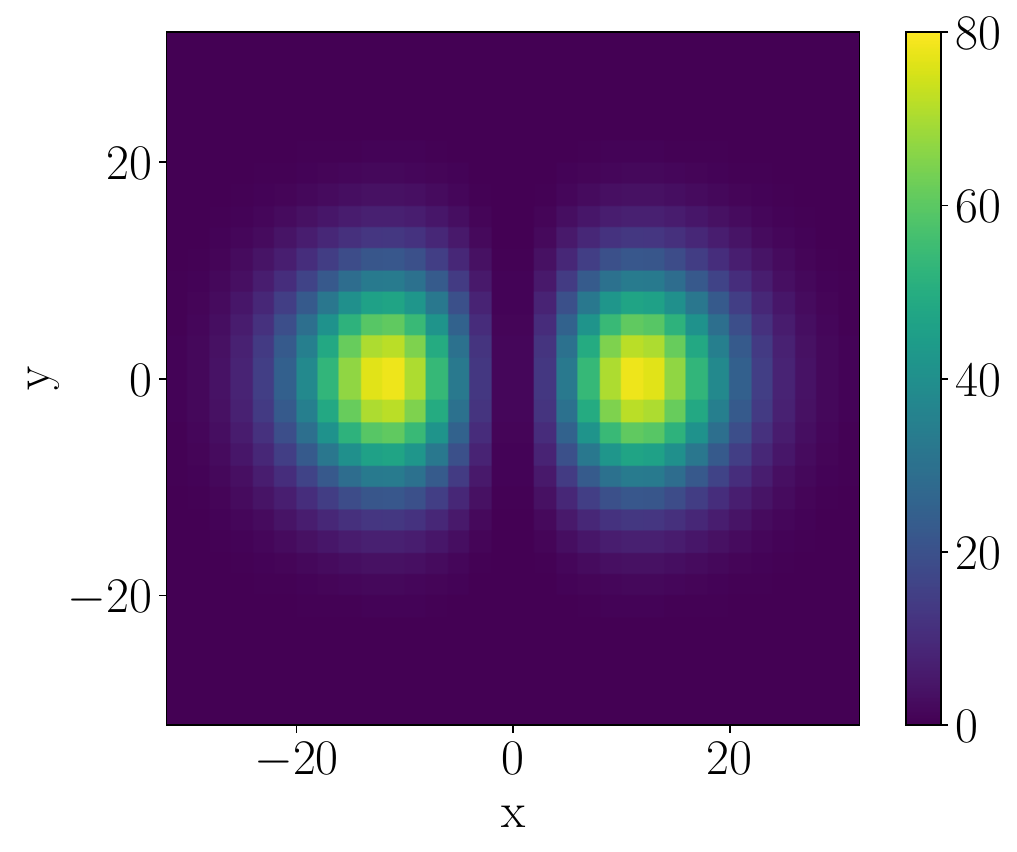}
}
    \caption{Square of the difference between distributions at reference angle $\vartheta_0=0^\circ$. Constant parameters were $\sigma=10$ and $\mu=10$. The grid size was $32\times32$ and $\Delta x=2$. Simulated angles $\vartheta=0^\circ,15^\circ,45^\circ,90^\circ,135^\circ,180^\circ$.}
    \label{fig:nsimdiff}
\end{figure*}

We ran a series of computer simulations to test the behavior of the FND for the direction algorithm. The reference angle was $\vartheta=0$. The distributions used had standard deviation $\sigma=10$ and radial centroid distance $\mu=2$, except for Figure~\ref{fig:nsimdiff}. Presented here is a total of five FND simulations. We simulate the behavior of the FND as the number of events $n$ increases for a constant bin width $\Delta x$. We vary $n$ for each of the bin widths $\Delta x=2,4,8,16$. The black curve is the analytical expression for the CFND of the form in (\ref{eq:finalfitcurve}) and should not be mistaken for a fit. In practice, the algorithm uses the minimum of an absolute sine fit to find the direction. This fit involves a vertical offset which is not discussed or tested here, but should be added in application of the algorithm. We also simulated the $n\rightarrow\infty$ case by sampling a Gaussian distribution and performing the FND algorithm, as shown in Figure~\ref{fig:simninfty}. Figure~\ref{fig:nsimdiff} shows the square of the difference between two Gaussian distributions as the reference dataset rotates about the origin, showing the algorithm process. A single FND point is computed by integrating the entire space and taking the square root.

\begin{figure*}[h]
    \centering
    \subfigure[$n=10$]{
        \includegraphics[width=0.3\textwidth]{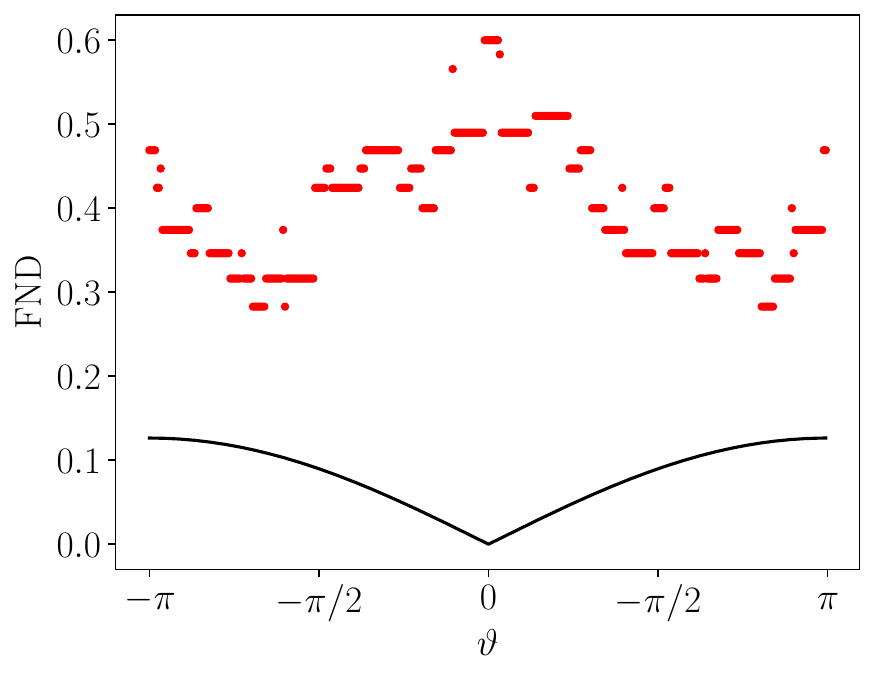}
    }
    \hfill
    \subfigure[$n=100$]{
        \includegraphics[width=0.3\textwidth]{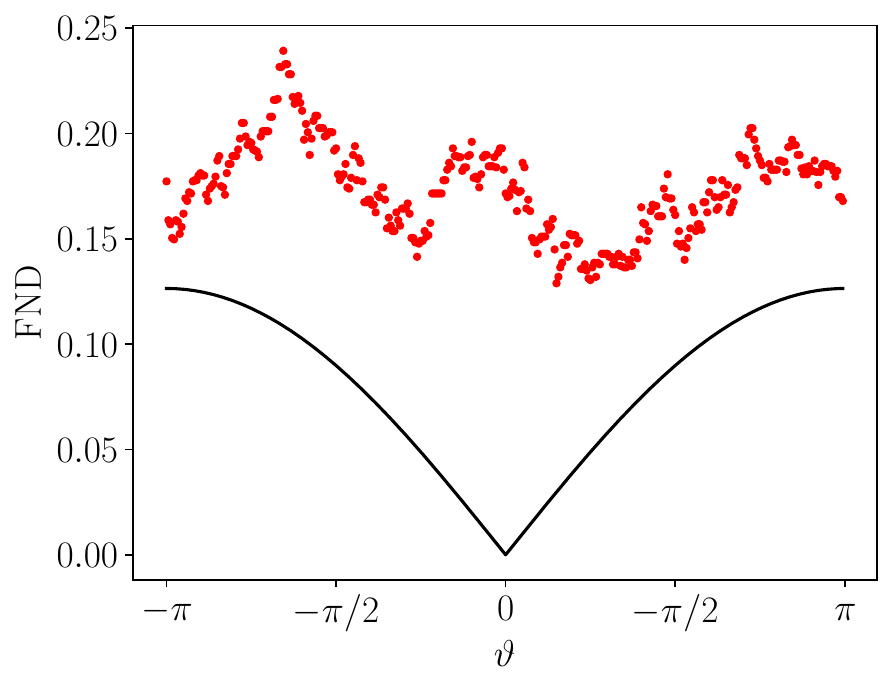}
    }
    \hfill
    \subfigure[$n=1000$]{
        \includegraphics[width=0.3\textwidth]{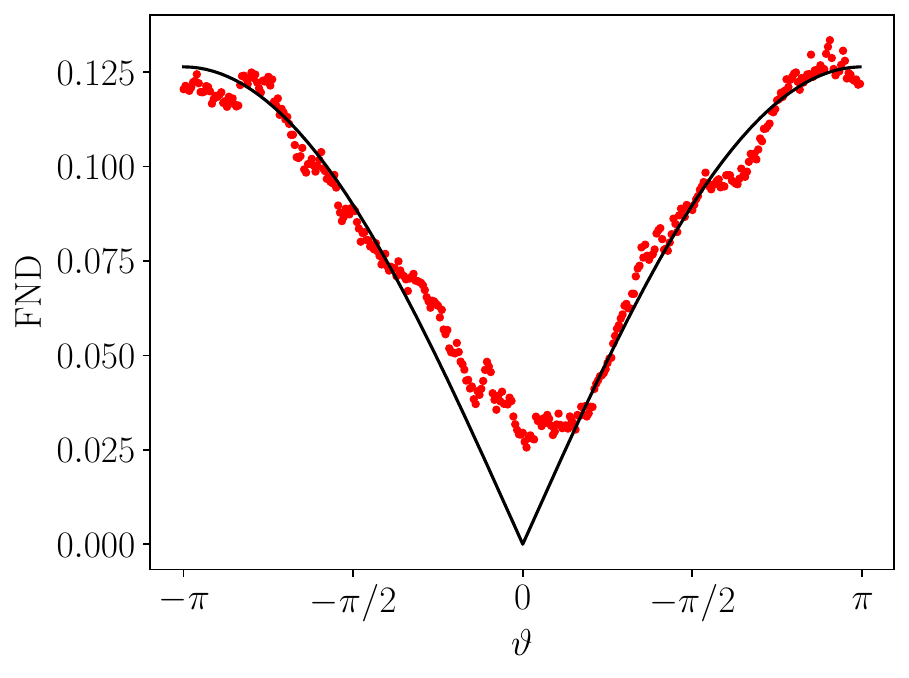}
    }
        \vspace{0.1em}
        \subfigure[$n=10^4$]{
        \includegraphics[width=0.3\textwidth]{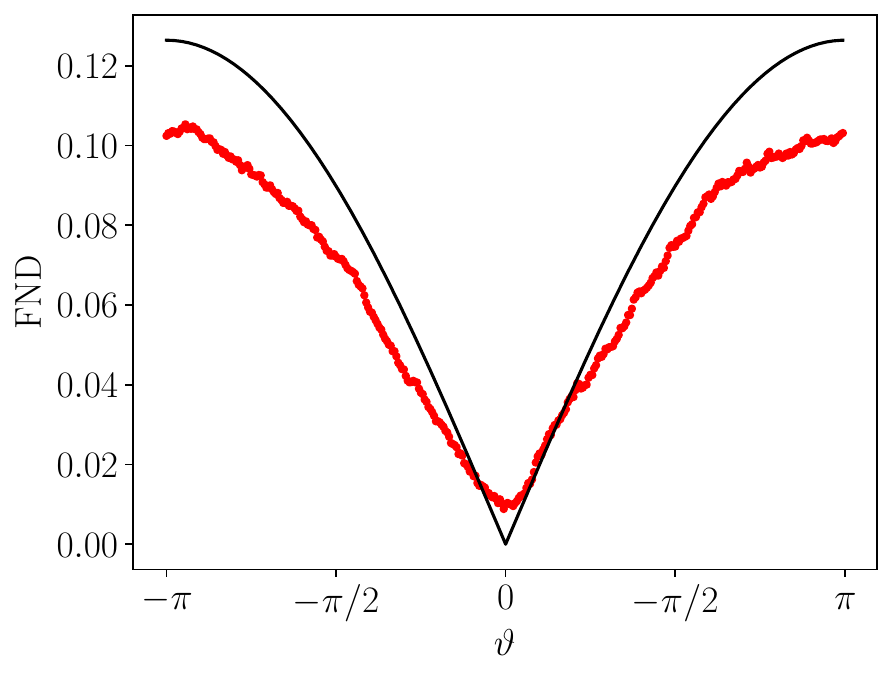}
    }
    \hfill
    \subfigure[$n=10^5$]{
        \includegraphics[width=0.3\textwidth]{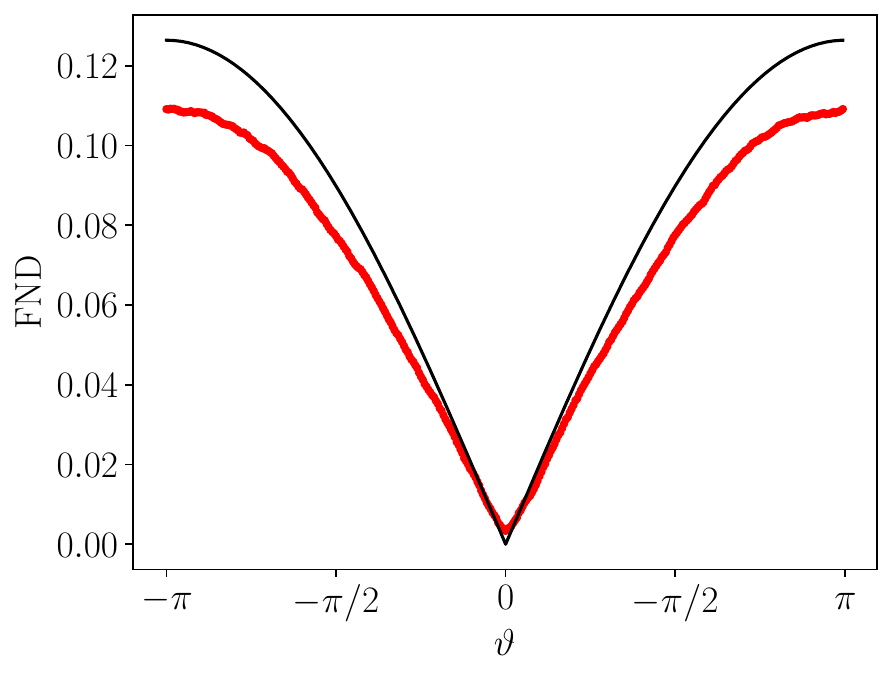}
    }
    \hfill
    \subfigure[$n=10^6$]{
        \includegraphics[width=0.3\textwidth]{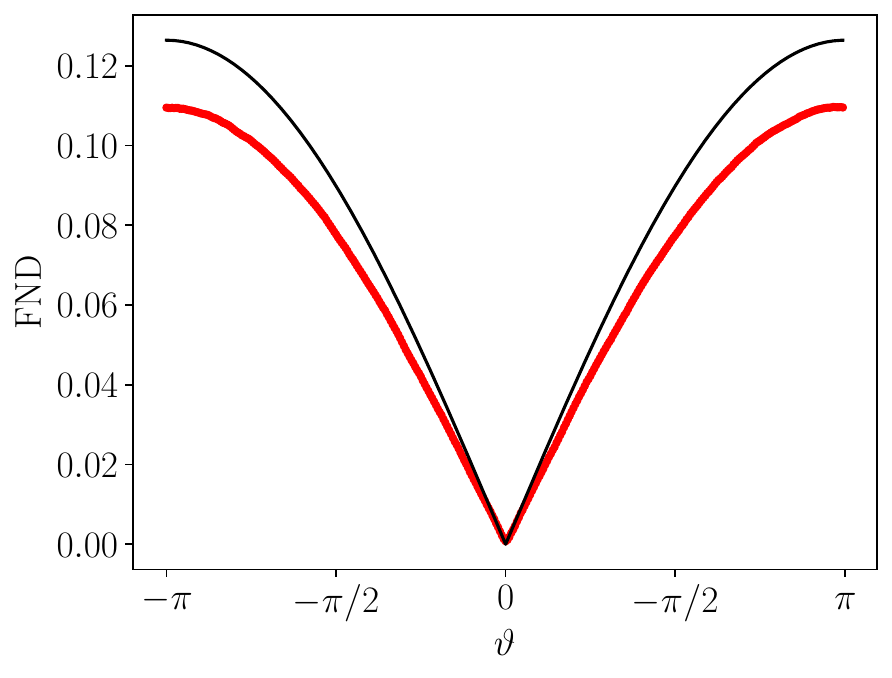}
}
    \caption{Simulation for increasing the total number of counts recorded in the histogram $n$. Other parameters stayed constant: $\sigma=10$, $\mu=2$, $\Delta x=16$. We ran six simulations for $n=10,100,1000,10^4,10^5,10^6$ events. The FND in red was simulated on an $8\times8$ histogram grid. The black curve is the absolute sine function in (\ref{eq:finalfitcurve}). Note that in this simulation all parameters are constant except for $n$ so the absolute sine curve is the same across all of these plots.}
    \label{fig:dx16}
\end{figure*}

\begin{figure*}[h]
    \centering
    \subfigure[$n=10$]{
        \includegraphics[width=0.3\textwidth]{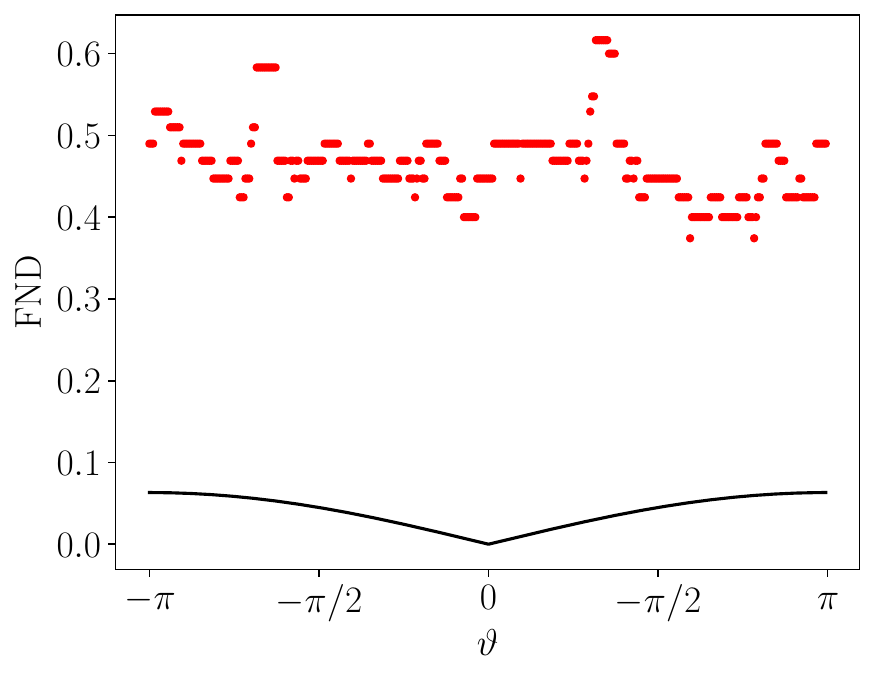}
    }
    \hfill
    \subfigure[$n=100$]{
        \includegraphics[width=0.3\textwidth]{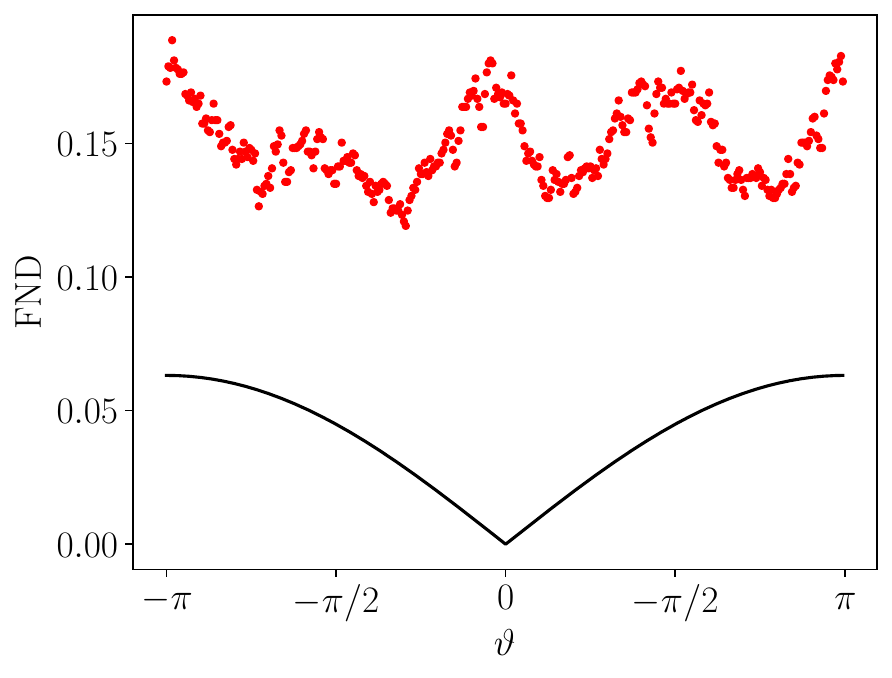}
    }
    \hfill
    \subfigure[$n=1000$]{
        \includegraphics[width=0.3\textwidth]{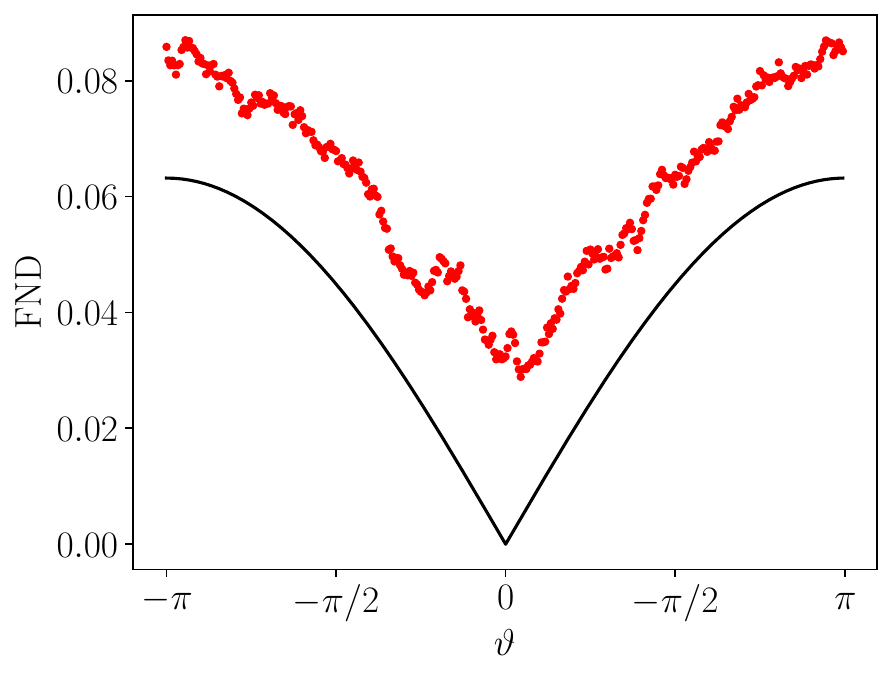}
    }
        \vspace{0.1em}
        \subfigure[$n=10^4$]{
        \includegraphics[width=0.3\textwidth]{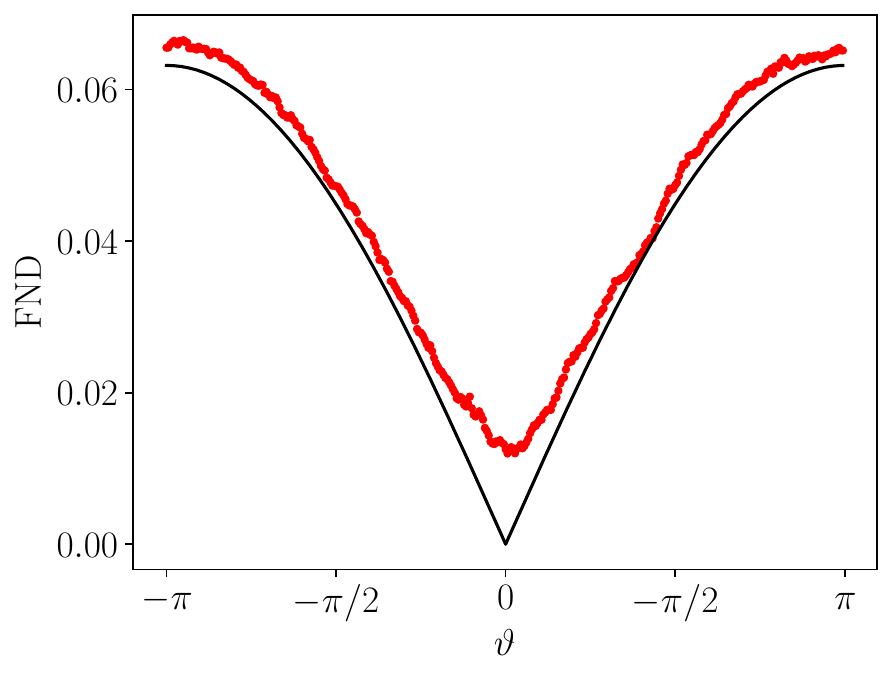}
    }
    \hfill
    \subfigure[$n=10^5$]{
        \includegraphics[width=0.3\textwidth]{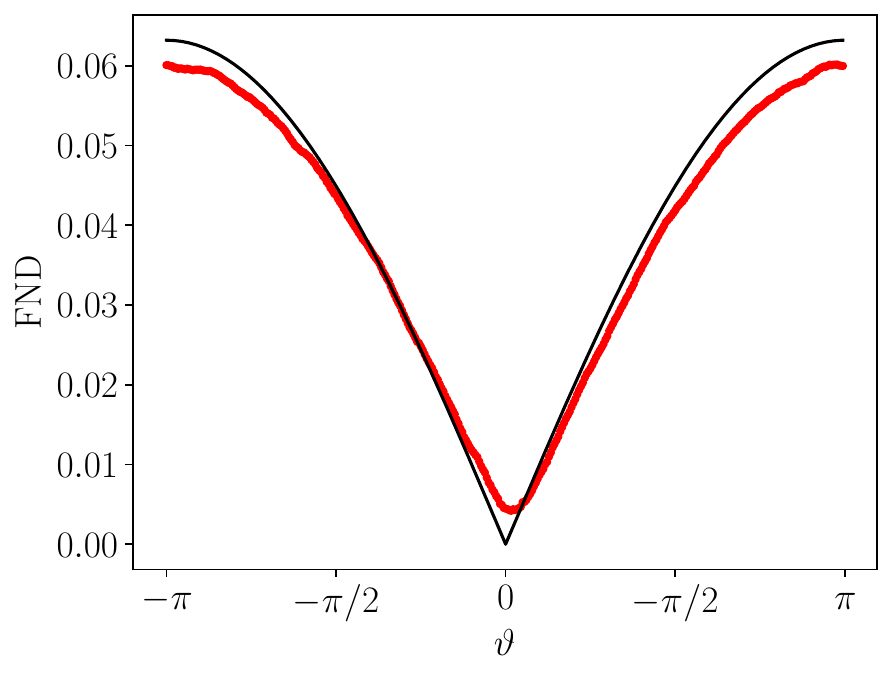}
    }
    \hfill
    \subfigure[$n=10^6$]{
        \includegraphics[width=0.3\textwidth]{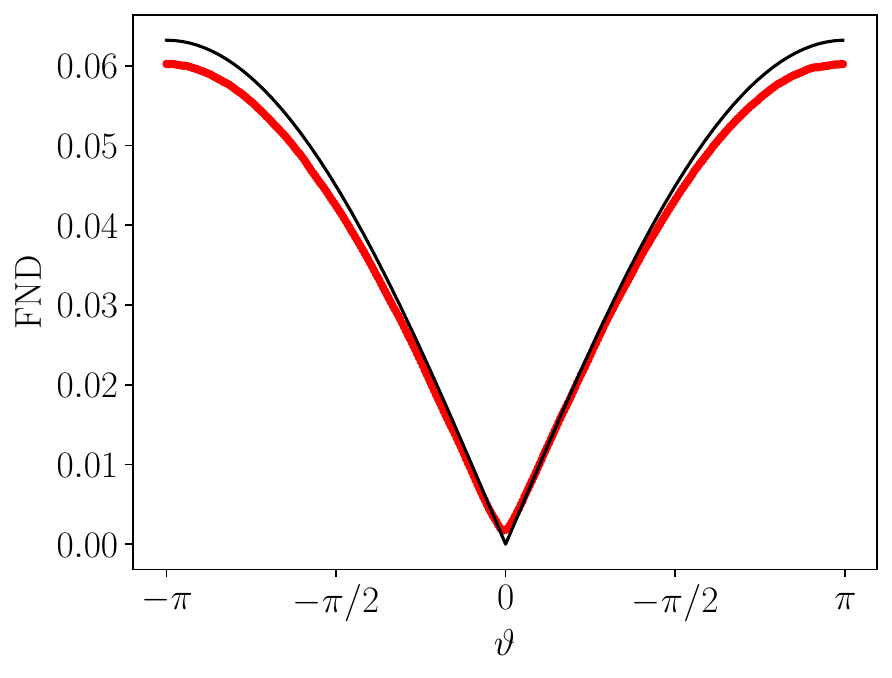}
}
    \caption{The parameters were $\sigma=10$, $\mu=2$, $\Delta x=8$. The size of the histogram grid was $16\times16$.}
    \label{fig:dx8}
\end{figure*}

\begin{figure*}[h]
    \centering
    \subfigure[$n=10$]{
        \includegraphics[width=0.3\textwidth]{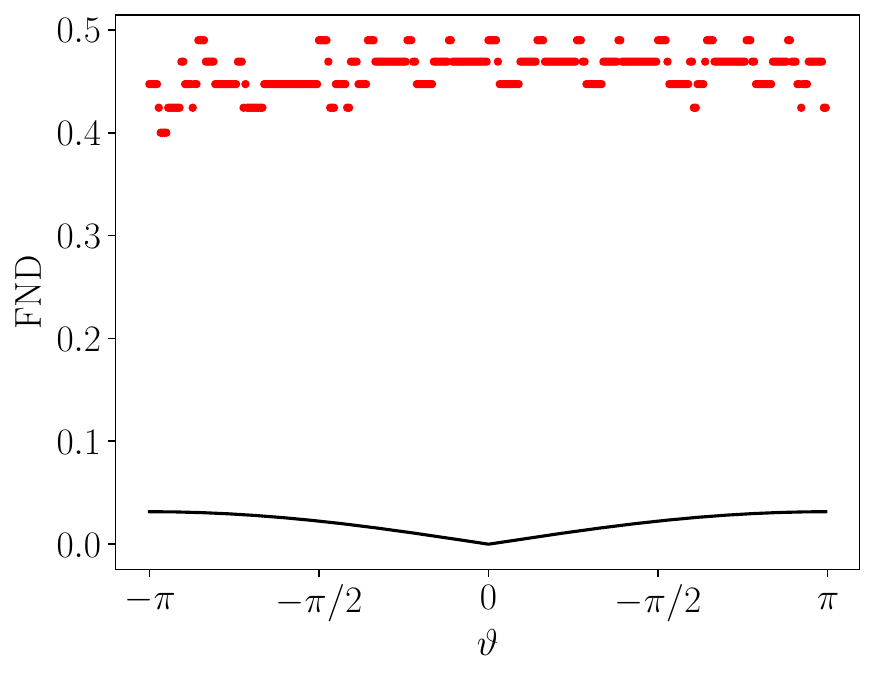}
    }
    \hfill
    \subfigure[$n=100$]{
        \includegraphics[width=0.3\textwidth]{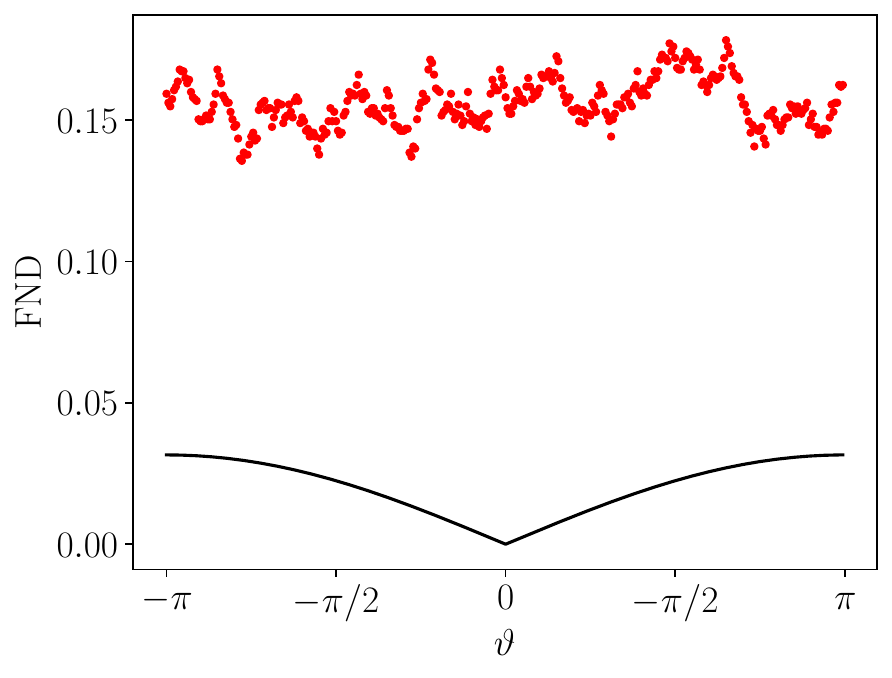}
    }
    \hfill
    \subfigure[$n=1000$]{
        \includegraphics[width=0.3\textwidth]{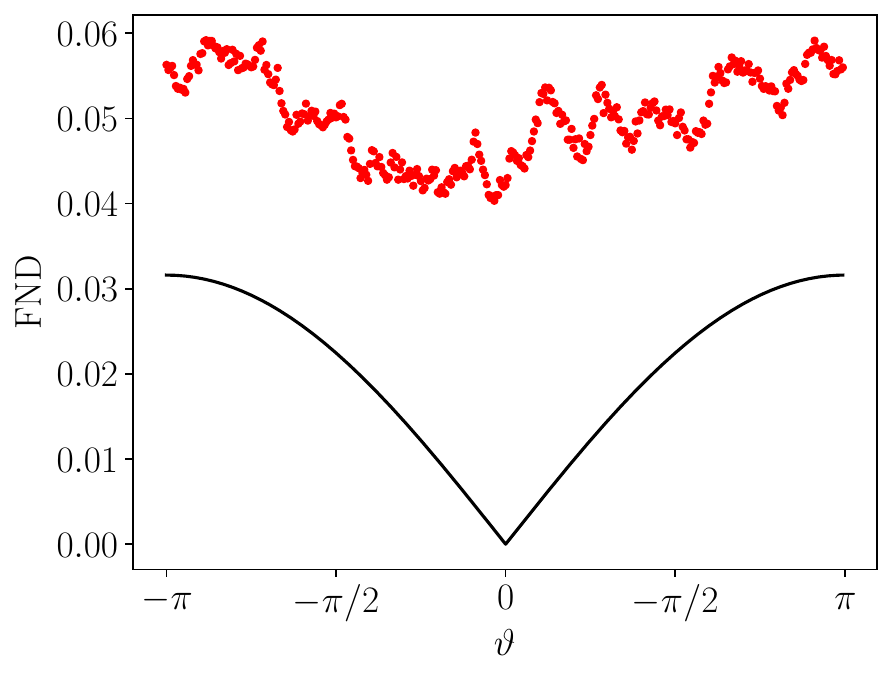}
    }
        \vspace{0.1em}
        \subfigure[$n=10^4$]{
        \includegraphics[width=0.3\textwidth]{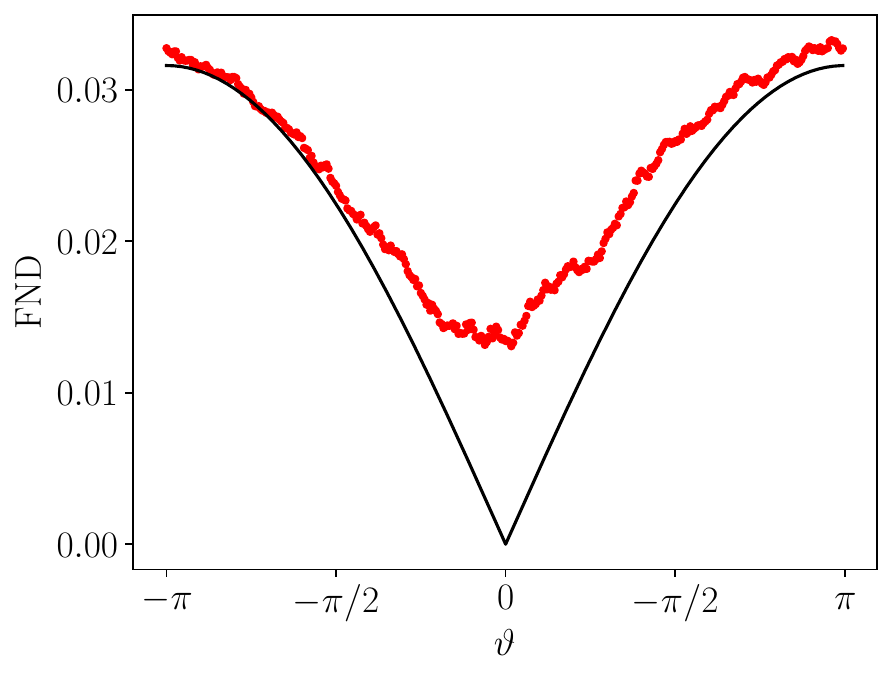}
    }
    \hfill
    \subfigure[$n=10^5$]{
        \includegraphics[width=0.3\textwidth]{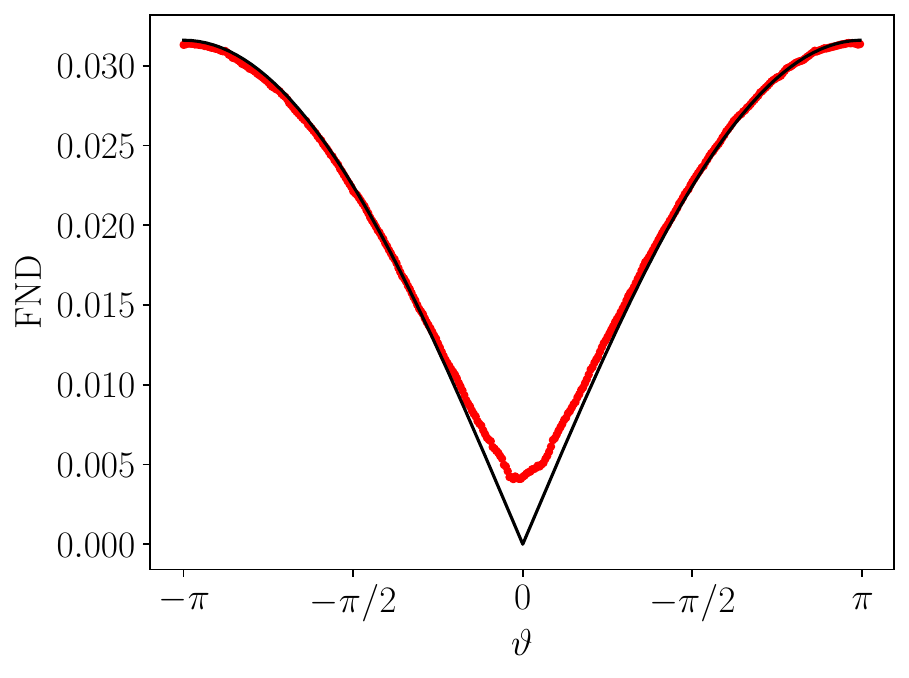}
    }
    \hfill
    \subfigure[$n=10^6$]{
        \includegraphics[width=0.3\textwidth]{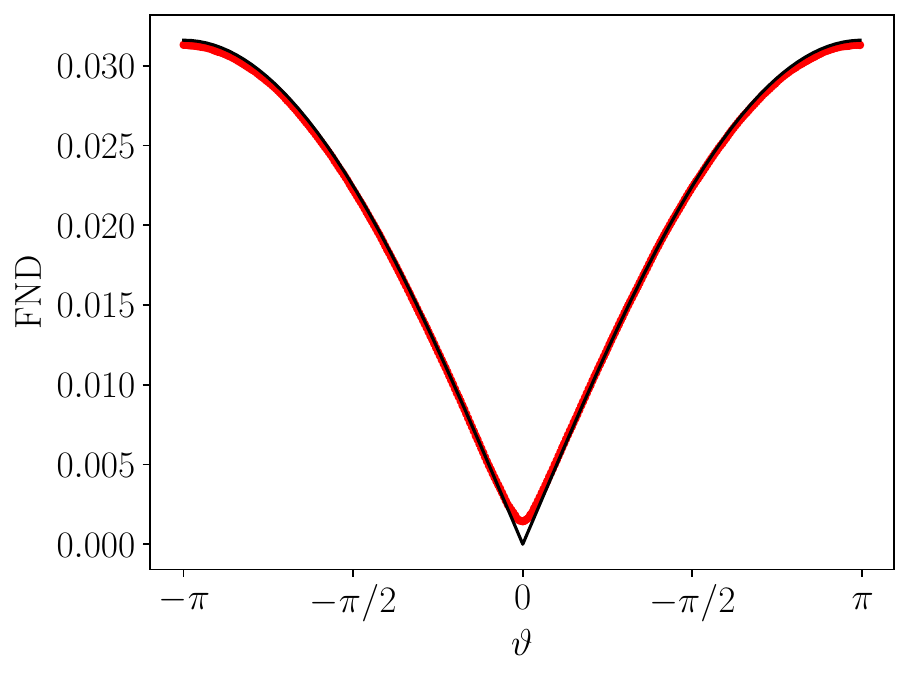}
}
    \caption{The parameters were $\sigma=10$, $\mu=2$, $\Delta x=4$. The size of the histogram grid was $32\times32$.}
    \label{fig:dx4}
\end{figure*}

\begin{figure*}[h]
    \centering
    \subfigure[$n=10$]{
        \includegraphics[width=0.3\textwidth]{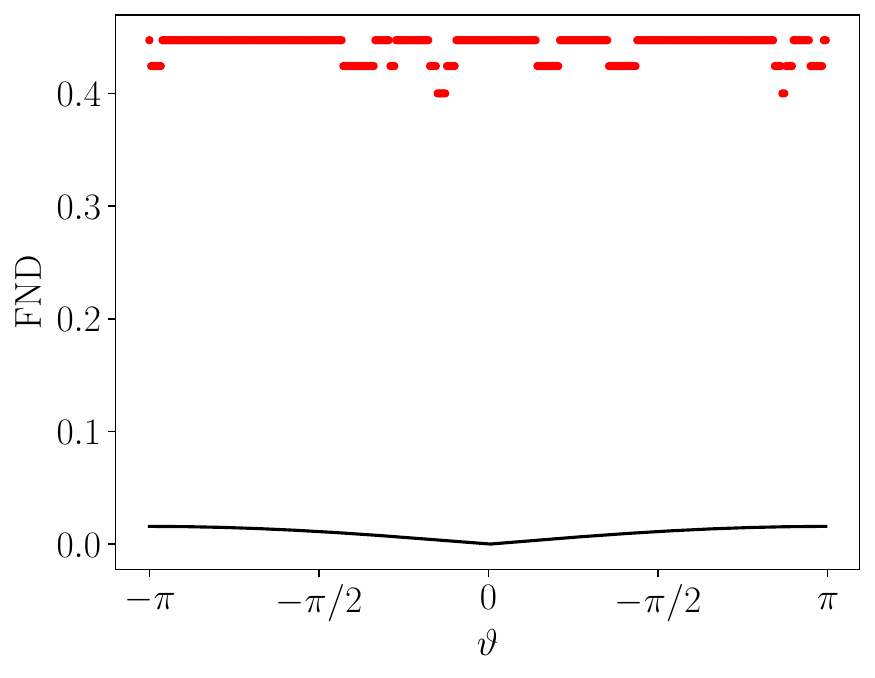}
    }
    \hfill
    \subfigure[$n=100$]{
        \includegraphics[width=0.3\textwidth]{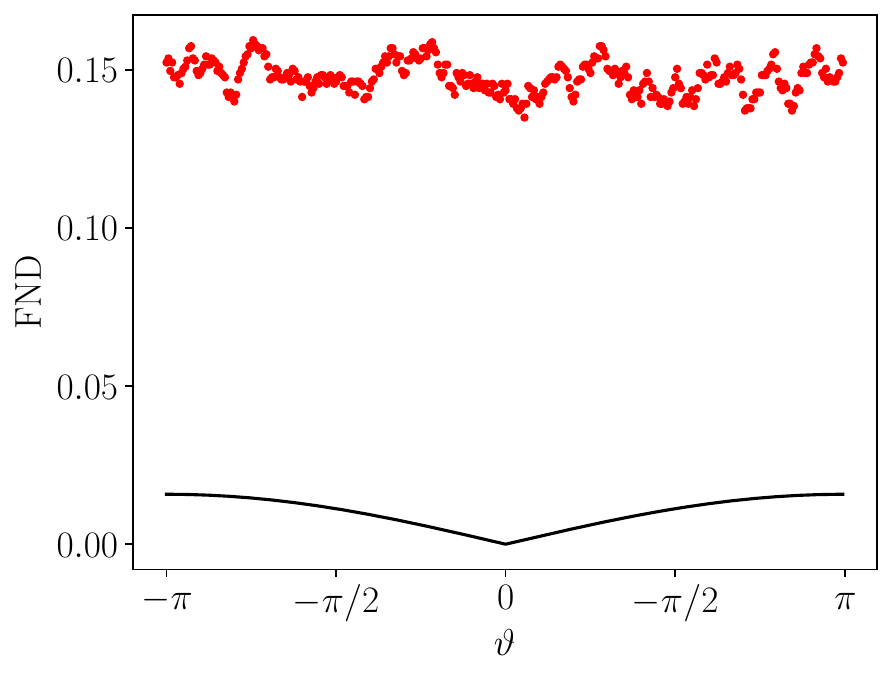}
    }
    \hfill
    \subfigure[$n=1000$]{
        \includegraphics[width=0.3\textwidth]{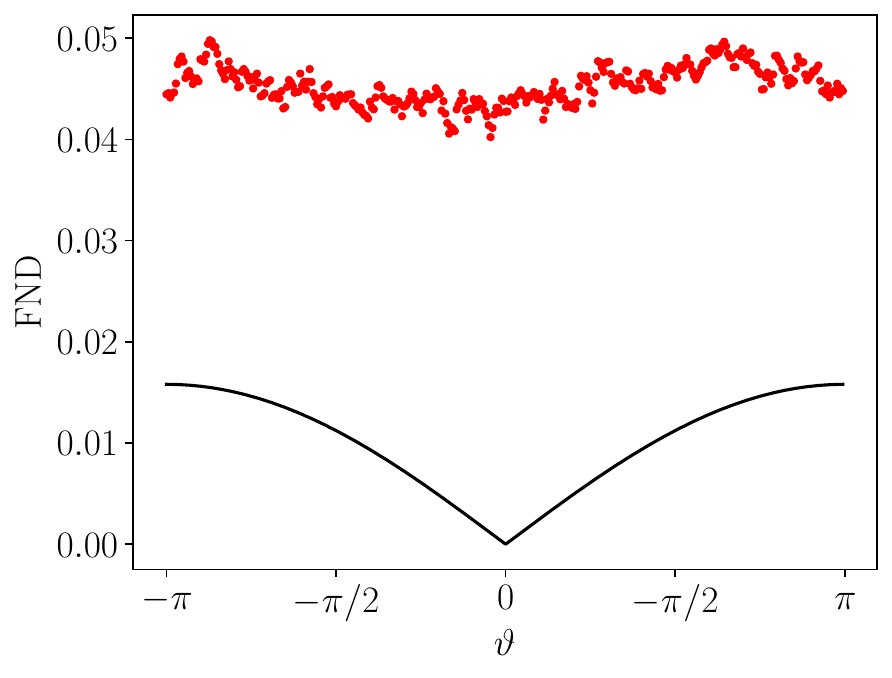}
    }
        \vspace{0.1em}
        \subfigure[$n=10^4$]{
        \includegraphics[width=0.3\textwidth]{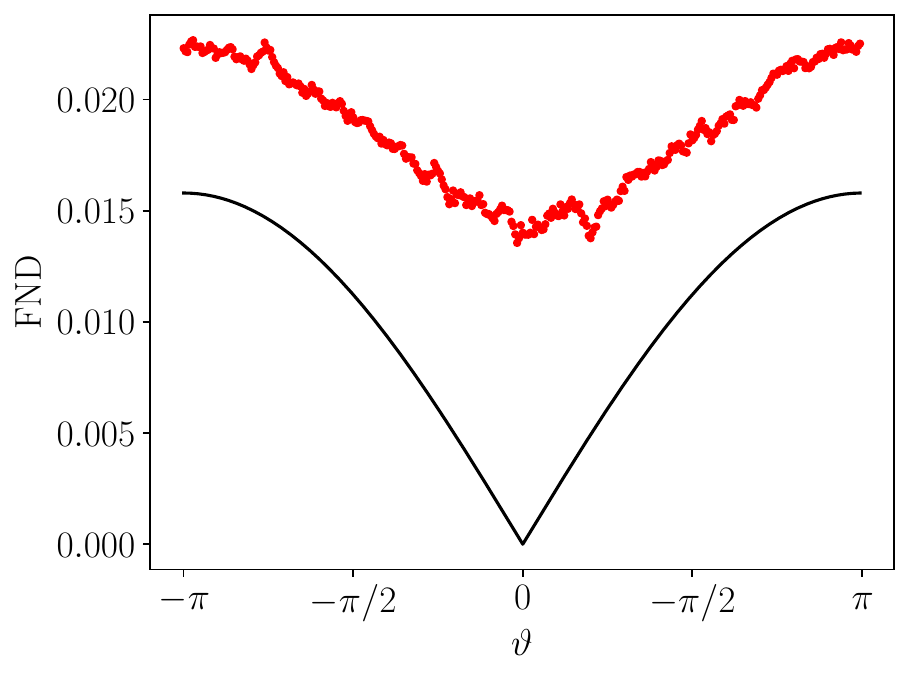}
    }
    \hfill
    \subfigure[$n=10^5$]{
        \includegraphics[width=0.3\textwidth]{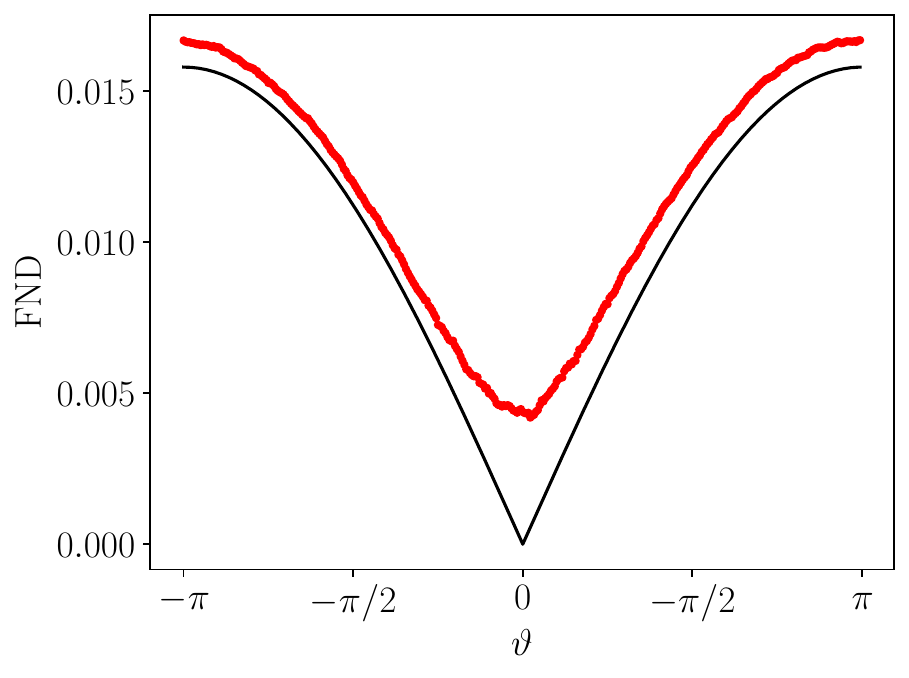}
    }
    \hfill
    \subfigure[$n=10^6$]{
        \includegraphics[width=0.3\textwidth]{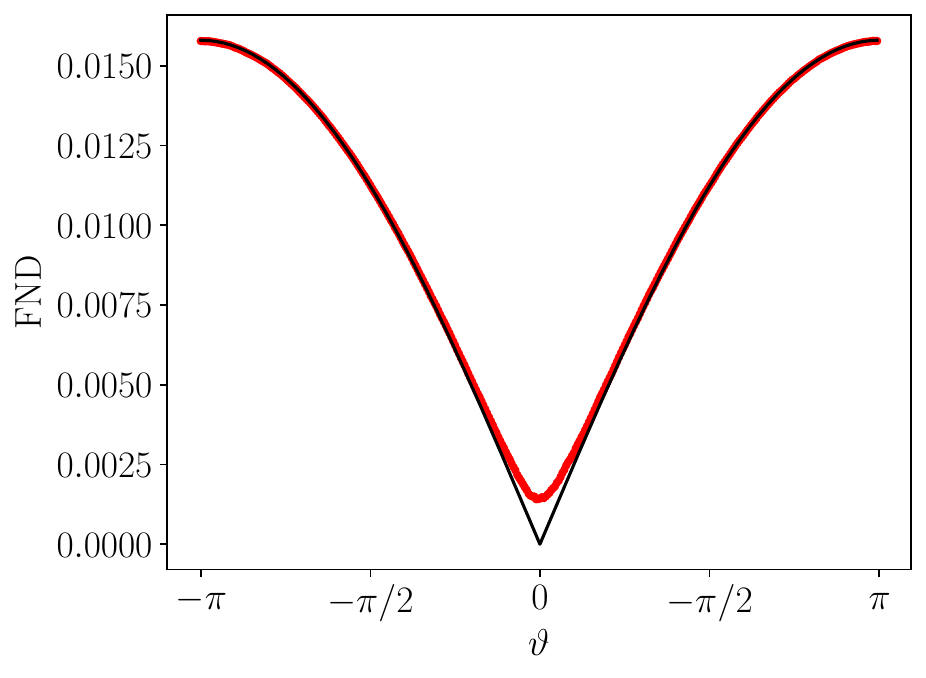}
}
    \caption{The parameters were $\sigma=10$, $\mu=2$, $\Delta x=2$. The size of the histogram grid was $64\times64$.}
    \label{fig:dx2}
\end{figure*}

\end{document}